\documentclass[10pt,twocolumn,journal]{IEEEtran}
\usepackage[T1]{fontenc}
\usepackage[latin9]{inputenc}
\usepackage{color}
\usepackage{float}
\usepackage{dsfont}
\usepackage{amsmath}
\usepackage{amssymb}
\usepackage{stackrel}
\usepackage{graphicx}
\usepackage{setspace}
\usepackage[unicode=true,
 bookmarks=true,bookmarksnumbered=true,bookmarksopen=true,bookmarksopenlevel=1,
 breaklinks=false,pdfborder={0 0 0},pdfborderstyle={},backref=false,colorlinks=false]
 {hyperref}
\hypersetup{pdftitle={Your Title},
 pdfauthor={Your Name},
 pdfpagelayout=OneColumn, pdfnewwindow=true, pdfstartview=XYZ, plainpages=false}

\makeatletter

\providecommand{\tabularnewline}{\\}
\floatstyle{ruled}
\newfloat{algorithm}{tbp}{loa}
\providecommand{\algorithmname}{Algorithm}
\floatname{algorithm}{\protect\algorithmname}

\usepackage[caption=false,font=footnotesize]{subfig}

\usepackage{algorithmic}
\usepackage{cite}

\makeatother

\begin{document}
\title{Joint Scattering Environment Sensing and Channel Estimation Based
on Non-stationary Markov Random Field}
\author{Wenkang Xu, Yongbo Xiao, An Liu,~\IEEEmembership{Senior Member,~IEEE,}
Ming Lei and Minjian Zhao{\normalsize{}}\thanks{Wenkang Xu, Yongbo Xiao, An Liu, Ming Lei, and Minjian Zhao are with
the College of Information Science and Electronic Engineering, Zhejiang
University, Hangzhou 310027, China (email: anliu@zju.edu.cn).}}
\maketitle
\begin{abstract}
This paper considers an integrated sensing and communication system,
where some radar targets also serve as communication scatterers. A
location domain channel modeling method is proposed based on the position
of targets and scatterers in the scattering environment, and the resulting
radar and communication channels exhibit a two-dimensional (2-D) joint
burst sparsity. We propose a joint scattering environment sensing
and channel estimation scheme to enhance the target/scatterer localization
and channel estimation performance simultaneously, where a spatially
non-stationary Markov random field (MRF) model is proposed to capture
the 2-D joint burst sparsity. An expectation maximization (EM) based
method is designed to solve the joint estimation problem, where the
E-step obtains the Bayesian estimation of the radar and communication
channels and the M-step automatically learns the dynamic position
grid and prior parameters in the MRF. However, the existing sparse
Bayesian inference methods used in the E-step involve a high-complexity
matrix inverse per iteration. Moreover, due to the complicated non-stationary
MRF prior, the complexity of M-step is exponentially large. To address
these difficulties, we propose an inverse-free variational Bayesian
inference algorithm for the E-step and a low-complexity method based
on pseudo-likelihood approximation for the M-step. In the simulations,
the proposed scheme can achieve a better performance than the state-of-the-art
method while reducing the computational overhead significantly.
\end{abstract}

\begin{IEEEkeywords}
Integrated sensing and communication, scattering environment sensing,
channel estimation, inverse-free, non-stationary Markov random field.

\thispagestyle{empty}
\end{IEEEkeywords}

\section{Introduction}

Radar sensing and wireless communication systems have been developed
independently for decades, and they are usually designed separately.
However, there are many similarities between sensing and communication
systems, such as signal processing algorithms, hardware architecture
and channel characteristics \cite{LiuFan_survey_signal_process,LiuFan_survey_dual_function,LiuFan_JRC1,LiuAn_survey_fundamental_limits}.
On the other hand, future communication signals will be able to support
high-accurate and robust sensing applications due to higher frequency
bands and larger antenna arrays \cite{Guerra_position,Sha_Position}.
Therefore, it is desirable to merge the sensing and communication
functionalities into a single system and jointly design the two functionalities
to meet high-performance sensing and communication requirements simultaneously.
In short, the sensing and communication functionalities are expected
to mutually assist each other by leveraging their similarities. 

We focus on an important property of the scattering environment in
massive multi-input multi-output (MIMO) Orthogonal Frequency Division
Multiplexing (OFDM) integrated sensing and communication (ISAC) systems,
which reflects an interesting similarity between radar sensing and
communication in terms of channel characteristics. The scattering
environment includes two subsets, i.e., radar targets and communication
scatterers, which contribute to the radar channel and communication
channel, respectively. However, some radar targets also serve as communication
scatterers in many cases. In an ISAC scenario for vehicle networks,
for instance, the BS needs to localize vehicles and obstacles on the
road and broadcast the sensing data to every vehicle to realize automatic
obstacle avoidance and route planning \cite{Zheng_autonomous_driving,Bilik_autonomous_driving}.
In this case, some vehicles and obstacles also contribute to communication
paths for neighboring vehicles. Recent literature has also concerned
this property of the scattering environment. In \cite{LiuFan_JRC1},
communication scatterers were assumed to be a subset of radar targets,
and thus the angle-of-arrivals (AoAs) of the communication channel
were also a subset of those of the radar channel. In \cite{Huangzhe_JRC2},
the authors assumed that radar targets and communication scatterers
partially overlapped, so the radar and communication channels shared
some common AoAs. Moreover, there are usually many different sizes
of scattering clusters in the scattering environment. Specifically,
if we treat a large target/scatterer as a cluster of point targets/scatterers,
then radar targets and communication scatterers can be viewed as scattering
clusters of different sizes. Therefore, the non-zeros elements of
sparse domain channels will appear in bursts \cite{Berger_CE_LASSO}.
Motivated by these, we want to exploit the important property of the
scattering environment to enhance both radar sensing and channel estimation
performance. We summarize some related works below.

\textbf{Joint target sensing and channel estimation:} In \cite{LiuFan_JRC1},
based on the assumption that targets also served as scatterers for
the communication signal, the authors proposed a novel target sensing
and channel estimation scheme. However, the target sensing and channel
estimation were carried out independently. In \cite{Huangzhe_JRC2},
the authors merged target sensing and channel estimation into a single
procedure under the assumption that radar targets and communication
scatterers partially overlapped. The authors in \cite{Wan_TSCE_UAV}
studied an application of ISAC for unmanned aerial vehicle (UAV) networks,
in which a UAV communicated with the terrestrial station while other
UAVs and obstacles were viewed as radar targets. A compressed sensing
based algorithm was designed to perform joint channel estimation and
target sensing to avoid UAV collisions. In \cite{Gaudio_TSCE_OTFS1,Gaudio_TSCE_OTFS2},
each radar target was also a communication receiver, and a two-step
approach was proposed to estimate the target location and the line-of-sight
(LoS) channel path.

\textbf{Joint scatterer/user localization and channel estimation:}
In \cite{Zhou_JLCE_position_modeling}, a massive MIMO-OFDM channel
was modeled based on the position of scatterers and a user, and then
the user location and channel coefficients were simultaneously estimated.
In \cite{LiuAn_JLCE_grid,LiuAn_directloc_vehicles,LiuGaunying_directloc_VBI,LiuAn_CE_Turbo_VBI},
a dynamic grid-based method was proposed to improve user localization
and channel estimation performance. In \cite{Yang_JLCE_soft_information},
the authors proposed to provide soft information about channel estimation
and user location instead of hard information about those. In \cite{Hong_Scatterer_Loc},
two geometry-based models were proposed for performing joint channel
estimation and scatterer localization involved in different bouncing
order propagation paths.

In this paper, we consider a broadband massive MIMO-OFDM ISAC system,
where the scattering environment sensing and channel estimation are
performed jointly to improve each other's performance. Here ``scattering
environment sensing'' refers to the localization of radar targets
and communication scatterers, and ``channel estimation'' refers
to the estimates of radar and communication channels. We notice that
the related work in \cite{Huangzhe_JRC2} considered a narrow-band
MIMO ISAC system and exploited the joint burst sparsity of the angular
domain channels to enhance both radar sensing and communication performance.
However, there are some new challenges when extending this work to
broadband MIMO-OFDM ISAC systems. First, the radar and communication
channels only share some common AoAs but not delay. Therefore, the
delay domain channels will no longer exhibit the joint burst sparsity.
Second, the hidden Markov model (HMM) used in \cite{Huangzhe_JRC2}
can only handle the one-dimensional burst sparsity but not high-dimensional
burst sparsity. Third, since the proposed turbo sparse Bayesian inference
(Turbo-SBI) algorithm involves the matrix inverse operation in each
iteration, it is very time-consuming when the problem size is large.

To address these difficulties, we improve our work in terms of channel
modeling method, sparse prior model, and algorithm design. A new joint
scattering environment sensing and channel estimation scheme is proposed.
Specifically, we first introduce the location domain channel modeling
method based on the assumption that part of radar targets and communication
scatterers share common positions. In this case, the resulting location
domain channels exhibit a two-dimensional (2-D) joint burst sparsity
naturally, as shown in Fig. \ref{fig:Illustration-of-radar}. Next,
we propose a non-stationary Markov random field (MRF) model, which
is able to deal with high-dimensional sparse structures with random
bursts. Finally, a new turbo inverse-free variational Bayesian inference
(Turbo-IF-VBI) algorithm is designed to reduce the computational complexity.
The main contributions are summarized below.

\textbf{A 2-D non-stationary Markov random field model \cite{book_MRF,MRF1,MRF2}:}
We propose a 2-D non-stationary Markov random field (MRF) model to
capture the 2-D joint burst sparsity of the location domain radar
and communication channels. The spatially non-stationary MRF model
has the flexibility to describe different degrees of sparsity and
different sizes of clusters, and therefore it can adapt to different
scattering environments that occur in practice.

\textbf{Turbo-IF-VBI algorithm:} The problem of joint scattering environment
sensing and channel estimation is formulated as a sparse Bayesian
inference (SBI) problem. Conventional sparse Bayesian inference algorithms,
such as the turbo variational Bayesian inference (Turbo-VBI) \cite{LiuAn_CE_Turbo_VBI}
and Turbo-SBI \cite{Huangzhe_JRC2} algorithms, involve a matrix inverse
in each iteration. Inspired by an inverse-free sparse Bayesian learning
(IF-SBL) framework that avoids the matrix inverse via maximizing a
relaxed evidence lower bound (ELBO) \cite{Duan_IFSBL}, we propose
a Turbo-IF-VBI algorithm with low complexity. In contrast to the IF-SBL,
our proposed Turbo-IF-VBI algorithm applies a three-layer sparse prior
model, which has the flexibility to exploit different types of sparse
structures.

\textbf{A low-complexity method to learn MRF parameters:} The spatially
non-stationary MRF has many unknown parameters that cannot be efficiently
learned by the conventional EM method because the computational complexity
is exponentially large. To overcome this challenge, we proposed a
low-complexity method based on pseudo-likelihood approximation to
approximately learn MRF parameters.

The rest of the paper is organized as follows. In Section II, we present
the system model. In Section III, we introduce the three-layer sparse
prior model and the non-stationary MRF model to capture the 2-D joint
burst sparsity of the location domain channels. In Section IV, we
present the proposed Turbo-IF-VBI algorithm and show its advantage
in terms of computational complexity. Simulation results and conclusion
are given in Section V and VI, respectively.

\textit{Notations}: $\left(\cdot\right)^{-1}$, $\left(\cdot\right)^{T}$,
$\left(\cdot\right)^{H}$, $\textrm{tr}\left(\cdot\right)$, $\textrm{diag}\left(\cdot\right)$,
and $\textrm{vec}\left(\cdot\right)$ denote the inverse, transpose,
conjugate transpose, trace, diagonalization, and vectorization operations,
respectively. $\left\Vert \cdot\right\Vert $ is the $\ell_{2}$ norm
of the given vector, $\otimes$ means Kronecker product operator,
$\textrm{BlockDiag}\left(\cdot\right)$ is block diagonalization of
the given matrices, $\mathbb{E}\left\{ \cdot\right\} $ denotes statistical
expectation, and $\mathfrak{Re}\left\{ \cdot\right\} $ represents
the real part of the argument. For a set $\mathcal{N}$, $\left|\mathcal{N}\right|$
is its cardinality. $\boldsymbol{x}\triangleq\left[x_{n}\right]_{n\in\mathcal{N}}\in\mathbb{C}^{\left|\mathcal{N}\right|\times1}$
is a vector composed of elements indexed by $\mathcal{N}$. $\mathbf{X}\triangleq\left[\mathbf{X}_{n}\right]_{n\in\mathcal{N}}\in\mathbb{C}^{M\left|\mathcal{N}\right|\times N}$
is a matrix composed of matrices indexed by $\mathcal{N}$, where
$\mathbf{X}_{n}\in\mathbb{C}^{M\times N}$. $\mathcal{CN}\left(\boldsymbol{x};\boldsymbol{\mu},\mathbf{\Sigma}\right)$
means that the vector $\boldsymbol{x}$ has a complex Gaussian distribution
with mean $\boldsymbol{\mu}$ and covariance matrix $\mathbf{\Sigma}$.
$\textrm{Gamma}\left(x;a,b\right)$ means that the variable $x$ follows
a gamma distribution with shape parameter $a$ and rate parameter
$b$.

\section{System Model}

\subsection{System Architecture and Frame Structure}

\begin{figure}[t]
\begin{centering}
\includegraphics[width=1\columnwidth]{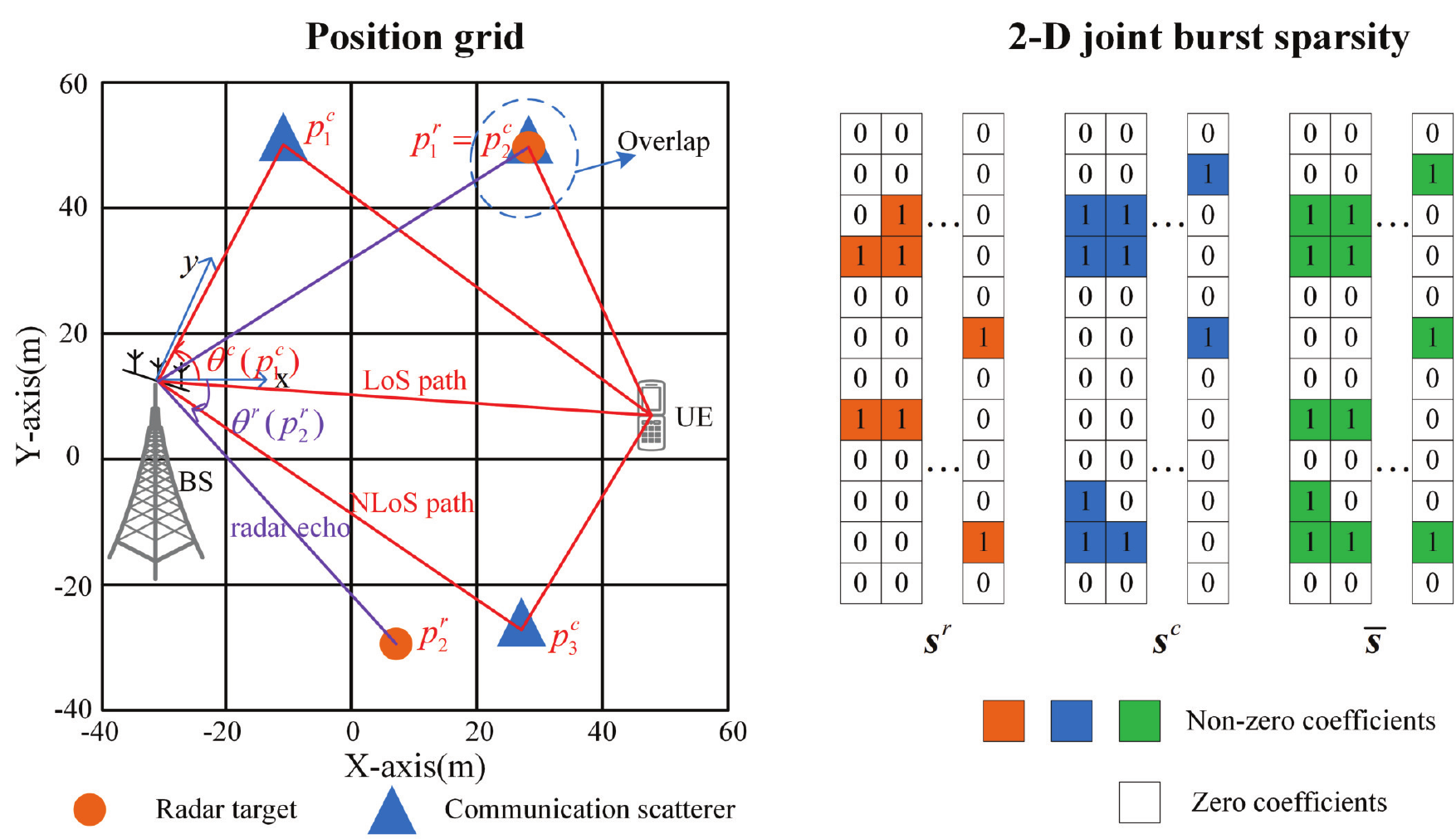}
\par\end{centering}
\caption{\label{fig:Illustration-of-radar}Illustration of location domain
radar and communication channels and their non-zero coefficients.}
\end{figure}

Consider a TDD massive MIMO-OFDM ISAC system, where one BS equipped
with $M\gg1$ antennas serves a single-antenna user while sensing
the scattering environment,\footnote{For clarity, we focus on the case with a single-antenna user system
in this paper. However, the proposed channel modeling method and signal
processing algorithm can be readily extended to the case with multiple
users by assigning orthogonal uplink pilots to different users.} as illustrated in Fig. \ref{fig:Illustration-of-radar}. The BS transmits
downlink pilots to sense the targets, and then the user transmits
uplink pilots to localize the scatterers and estimate the communication
channel. Suppose there are a total number of $K$ targets and $L$
communication scatterers in the scattering environment. As discussed
above, there might be some overlap between targets and communication
scatterers. The user is located at $\boldsymbol{p}_{u}=\left[p^{x},p^{y}\right]^{T}$
in a 2-D area $\mathcal{R}$. The BS is located at a known position
$\boldsymbol{p}_{b}=\left[\tilde{p}^{x},\tilde{p}^{y}\right]^{T}$.
Let $\boldsymbol{p}_{\mathrm{\mathit{k}}}^{r}=\left[p_{k}^{r,x},p_{k}^{r,y}\right]^{T}$
and $\boldsymbol{p}_{\mathrm{\mathit{l}}}^{c}=\left[p_{l}^{c,x},p_{l}^{c,y}\right]^{T}$
be the coordinates of the $k\textrm{-th}$ target and the $l\textrm{-th}$
communication scatterer, respectively. Moreover, we assume that the
BS has some prior information about the user location based on the
Global Positioning System (GPS) or the previous user localization
result \footnote{The knowledge of the transmitter location (i.e., the user location
in this case) is usually required for performing scatterer localization
\cite{LiuAn_directloc_vehicles,Hong_Scatterer_Loc}.}.

Note that we focus on a 2-D scenario in this paper that is suitable
for some application scenarios, such as high-way vehicle networks,
in which the mobile user, targets, and communication scatterers are
mainly located on the road. However, our proposed scheme can also
be easily extended to three-dimensional (3-D) scenarios by adding
the third dimension (the z coordinate axis) to the location domain.
Besides, the effect of clutters can be incorporated in the proposed
model and algorithm. Specifically, the weak clutters can be absorbed
into the noise, while the strong clutters can be treated as targets
of non-interest, whose parameters will also be estimated. After all
the targets have been detected, we can further identify the targets
of interest or non-interest based on the properties/features of their
parameters.
\begin{figure}[t]
\begin{centering}
\includegraphics[width=1\linewidth]{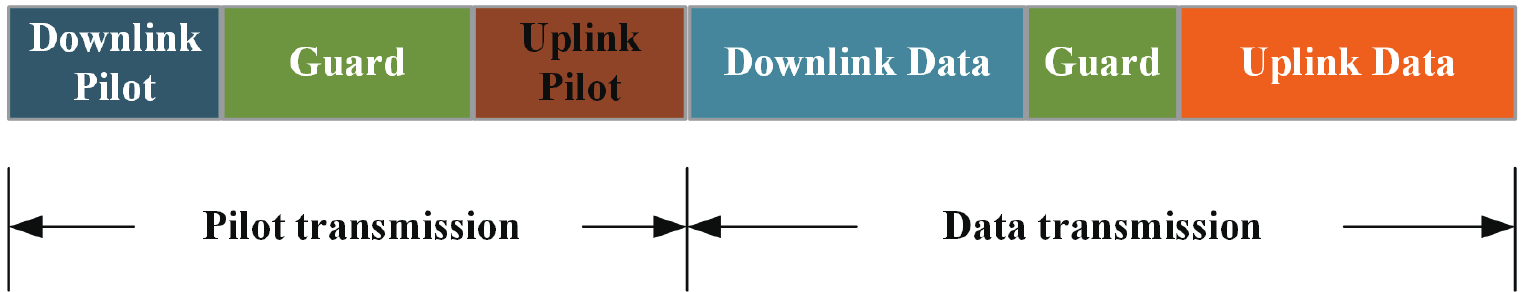}
\par\end{centering}
\caption{\label{fig:Frame-structure}Frame structure of the ISAC system.}
\end{figure}

The time is divided into frames, with each frame containing two phases:
the pilot transmission phase and the data transmission phase, as illustrated
in Fig. \ref{fig:Frame-structure}. We will focus on the pilot transmission
phase which combines the scattering environment sensing and channel
estimation into a single procedure. Specifically, the BS first periodically
scans broad angular sectors and transmits downlink pilots to sense
the targets at each angular sector. Then the user transmits uplink
pilots to the BS for channel estimation. If the user is in a certain
angular sector, there will be much overlap between the targets in
this angular sector and scatterers associated with this user. Finally,
for each angular sector, the BS performs joint scattering environment
sensing and channel estimation based on the reflected downlink pilot
signal toward this angular sector and the uplink pilot signal of the
user in this angular sector\footnote{Note that the BS can determine whether a user lies in a broad angular
sector by using the prior information about the user location.}. A guard interval is required between downlink and uplink pilots
to avoid interference. In the rest of the paper, we shall focus on
the problem of joint scattering environment sensing and channel estimation
for one angular sector.

\subsection{Reflected Downlink Pilot Signal}

Target sensing aims at detecting the presence of the target and estimating
the target location. To achieve this, on the $n\textrm{-th}$ subcarrier
for $n\in\mathcal{N}_{b}$, the BS transmits a downlink pilot $\boldsymbol{v}_{n}^{r}\in\mathbb{C}^{M\times1}$
toward the desired angular sector, and the received signal reflected
from the targets can be expressed as
\begin{equation}
\boldsymbol{y}_{n}^{r}=\mathbf{H}_{n}^{r}\boldsymbol{v}_{n}^{r}+\boldsymbol{z}_{n}^{r},\text{ }\forall n\in\mathcal{N}_{b},
\end{equation}
where $\mathbf{H}_{n}^{r}\in\mathbb{C}^{M\times M}$ denotes the radar
channel matrix, $\boldsymbol{z}_{n}^{r}\in\mathbb{C}^{M\times1}$
is the additive white Gaussian noise (AWGN) with variance $1/\gamma^{r}$,
and $\mathcal{N}_{b}$ is the set of subcarriers used for target sensing
in the desired angular sector. Let $\theta^{r}\left(\boldsymbol{p}_{k}^{r}\right)$
and $\tau^{r}\left(\boldsymbol{p}_{k}^{r}\right)$ represent the AoA
and delay of the $k\textrm{-th}$ target, respectively, which are
related to the position of the BS and the $k\textrm{-th}$ target
through
\begin{equation}
\begin{aligned}\theta^{r}\left(\boldsymbol{p}_{k}^{r}\right) & =\arctan\left(\frac{p_{k}^{r,y}-\tilde{p}^{y}}{p_{k}^{r,x}-\tilde{p}^{x}}\right)+\pi\cdotp\mathds{1}\left(p_{k}^{r,x}<\tilde{p}^{x}\right),\\
\tau^{r}\left(\boldsymbol{p}_{k}^{r}\right) & =2\left\Vert \boldsymbol{p}_{b}-\boldsymbol{p}_{k}^{r}\right\Vert /c,
\end{aligned}
\end{equation}
where the angle is calculated anticlockwise with respect to the \textit{x}-axis,
$\mathds{1}\left(\textrm{E}\right)=1$ if the logical expression $\textrm{E}$
is true, and $c$ denotes the speed of light. Then the radar channel
matrix can be modeled as
\begin{equation}
\mathbf{H}_{n}^{r}=\sum_{k=0}^{K}x_{k}^{r}e^{-j2\pi nf_{0}\left(\tau^{r}\left(\boldsymbol{p}_{k}^{r}\right)\right)}\boldsymbol{a}\left(\theta^{r}\left(\boldsymbol{p}_{k}^{r}\right)\right)\boldsymbol{a}^{T}\left(\theta^{r}\left(\boldsymbol{p}_{k}^{r}\right)\right),\label{eq:Hnr}
\end{equation}
where $x_{k}^{r}$ represents radar cross section of the $k\textrm{-th}$
target, $f_{0}$ is the subcarrier interval, and $\boldsymbol{a}\left(\theta\right)\in\mathbb{C}^{M\times1}$
denotes the array response vector at the BS. For the special case
of a uniform linear array (ULA), we have
\begin{equation}
\boldsymbol{a}\left(\theta\right)=\frac{1}{\sqrt{M}}\left[1,e^{j\pi\sin\theta},\ldots,e^{j\left(M-1\right)\pi\sin\theta}\right]^{T}.
\end{equation}
Note that in (\ref{eq:Hnr}), we treat the mobile user as the $0\textrm{-th}$
target with its position $\boldsymbol{p}_{0}^{r}\triangleq\boldsymbol{p}_{u}$.
If the BS can ``see'' the user through the radar echo signal, we
have $\left|x_{0}^{r}\right|>0$. In this case, the echo signal also
directly provides some additional information to assist in locating
the user's position. Otherwise, we have $x_{0}^{r}=0$.

\subsection{Received Uplink Pilot Signal}

The uplink pilot is used to estimate the communication channel as
well as sense the communication scatterers between the user and the
BS. On the $n\textrm{-th}$ subcarrier for $n\in\mathcal{N}_{u}$,
the user transmits an uplink pilot $u_{n}^{c}\in\mathbb{C}$ and then
the BS receives the signal, which can be expressed as
\begin{equation}
\boldsymbol{y}_{n}^{c}=\boldsymbol{h}_{n}^{c}u_{n}^{c}+\boldsymbol{z}_{n}^{c},\text{ }\forall n\in\mathcal{N}_{u}
\end{equation}
where $\boldsymbol{h}_{n}^{c}\in\mathbb{C}^{M\times1}$ denotes the
communication channel vector, $\boldsymbol{z}_{n}^{c}\in\mathbb{C}^{M\times1}$
is the the AWGN with variance $1/\gamma^{c}$, and $\mathcal{N}_{u}$
is the set of subcarriers used for uplink channel estimation for the
user.

Assume that there is one LoS path, $L$ single-bounce non-LoS (NLoS)
paths corresponding to the $L$ communication scatterers, and $J$
multiple-bounce NLoS paths for the communication channel \cite{Hong_Scatterer_Loc}.
Let $\theta^{c}\left(\boldsymbol{p}_{u}\right)$ and $\theta^{c}\left(\boldsymbol{p}_{l}^{c}\right)$
represent the AoAs of the LoS path and the $l\textrm{-th}$ single-bounce
NLoS path, respectively, which are related to the position of the
BS, the user, and the $l\textrm{-th}$ communication scatterer through
\begin{equation}
\begin{aligned}\theta^{c}\left(\boldsymbol{p}_{u}\right) & =\arctan\left(\frac{p^{y}-\tilde{p}^{y}}{p^{x}-\tilde{p}^{x}}\right)+\pi\cdotp\mathds{1}\left(p^{x}<\tilde{p}^{x}\right),\\
\theta^{c}\left(\boldsymbol{p}_{l}^{c}\right) & =\arctan\left(\frac{p_{l}^{c,y}-\tilde{p}^{y}}{p_{l}^{c,x}-\tilde{p}^{x}}\right)+\pi\cdotp\mathds{1}\left(p_{l}^{c,x}<\tilde{p}^{x}\right).
\end{aligned}
\end{equation}
Clearly, the relative delay of the $l\textrm{-th}$ single-bounce
NLoS path (relative to the LoS path) can be expressed as
\begin{equation}
\tau^{c}\left(\boldsymbol{p}_{l}^{c},\boldsymbol{p}_{u}\right)=\left(\left\Vert \boldsymbol{p}_{b}-\boldsymbol{p}_{l}^{c}\right\Vert +\left\Vert \boldsymbol{p}_{u}-\boldsymbol{p}_{l}^{c}\right\Vert -\left\Vert \boldsymbol{p}_{b}-\boldsymbol{p}_{u}\right\Vert \right)/c.
\end{equation}
Furthermore, let $\tilde{\theta}_{j}^{c}$ and $\tilde{\tau}_{j}^{c}$
denote the AoA and relative delay of the $j\textrm{-th}$ multiple-bounce
NLoS path, respectively.

Then the communication channel vector can be modeled as
\begin{equation}
\boldsymbol{h}_{n}^{c}=\boldsymbol{h}_{n}^{c,\textrm{L}}+\boldsymbol{h}_{n}^{c,\textrm{SL}}+\boldsymbol{h}_{n}^{c,\textrm{ML}},\label{eq:hnc}
\end{equation}
with
\begin{subequations}
\begin{align}
\boldsymbol{h}_{n}^{c,\textrm{L}} & =x_{0}^{c}e^{-j2\pi nf_{0}\tau_{o}}\boldsymbol{a}\left(\theta^{c}\left(\boldsymbol{p}_{u}\right)\right),\label{eq:hncL}\\
\boldsymbol{h}_{n}^{c,\textrm{SL}} & =\sum_{l=1}^{L}x_{l}^{c}e^{-j2\pi nf_{0}\left(\tau^{c}\left(\boldsymbol{p}_{l}^{c},\boldsymbol{p}_{u}\right)+\tau_{o}\right)}\boldsymbol{a}\left(\theta^{c}\left(\boldsymbol{p}_{l}^{c}\right)\right),\label{eq:hncNL}\\
\boldsymbol{h}_{n}^{c,\textrm{ML}} & =\sum_{j=1}^{J}\tilde{x}_{j}^{c}e^{-j2\pi nf_{0}\left(\tilde{\tau}_{j}^{c}+\tau_{o}\right)}\boldsymbol{a}\left(\tilde{\theta}_{j}^{c}\right),\label{eq:hncMNL}
\end{align}
\end{subequations}
where $\boldsymbol{h}_{n}^{c,\textrm{L}}$, $\boldsymbol{h}_{n}^{c,\textrm{SL}}$,
and $\boldsymbol{h}_{n}^{c,\textrm{ML}}$ represent the channel response
vectors of the LoS path, single-bounce NLoS paths, and multiple-bounce
NLoS paths, respectively, $x_{0}^{c}$, $x_{l}^{c}$, and $\tilde{x}_{j}^{c}$
denote the channel gains of the LoS path, the $l\textrm{-th}$ single-bounce
NLoS path, and the $j\textrm{-th}$ multiple-bounce NLoS path, respectively,
and $\tau_{o}$ is the time offset (relative to the LoS path) caused
by the timing synchronization error at the BS.

\section{Sparse Bayesian Inference Formulation}

In this section, we first obtain a sparse representation of the radar
and communication channels. Then, we introduce a three-layer sparse
prior and a spatially non-stationary MRF model to capture the 2-D
joint burst sparsity of the location domain channels. Finally, we
formulate the problem of joint scattering environment sensing and
channel estimation as a sparse Bayesian inference problem.
\begin{figure*}[t]
\begin{centering}
\includegraphics[width=120mm]{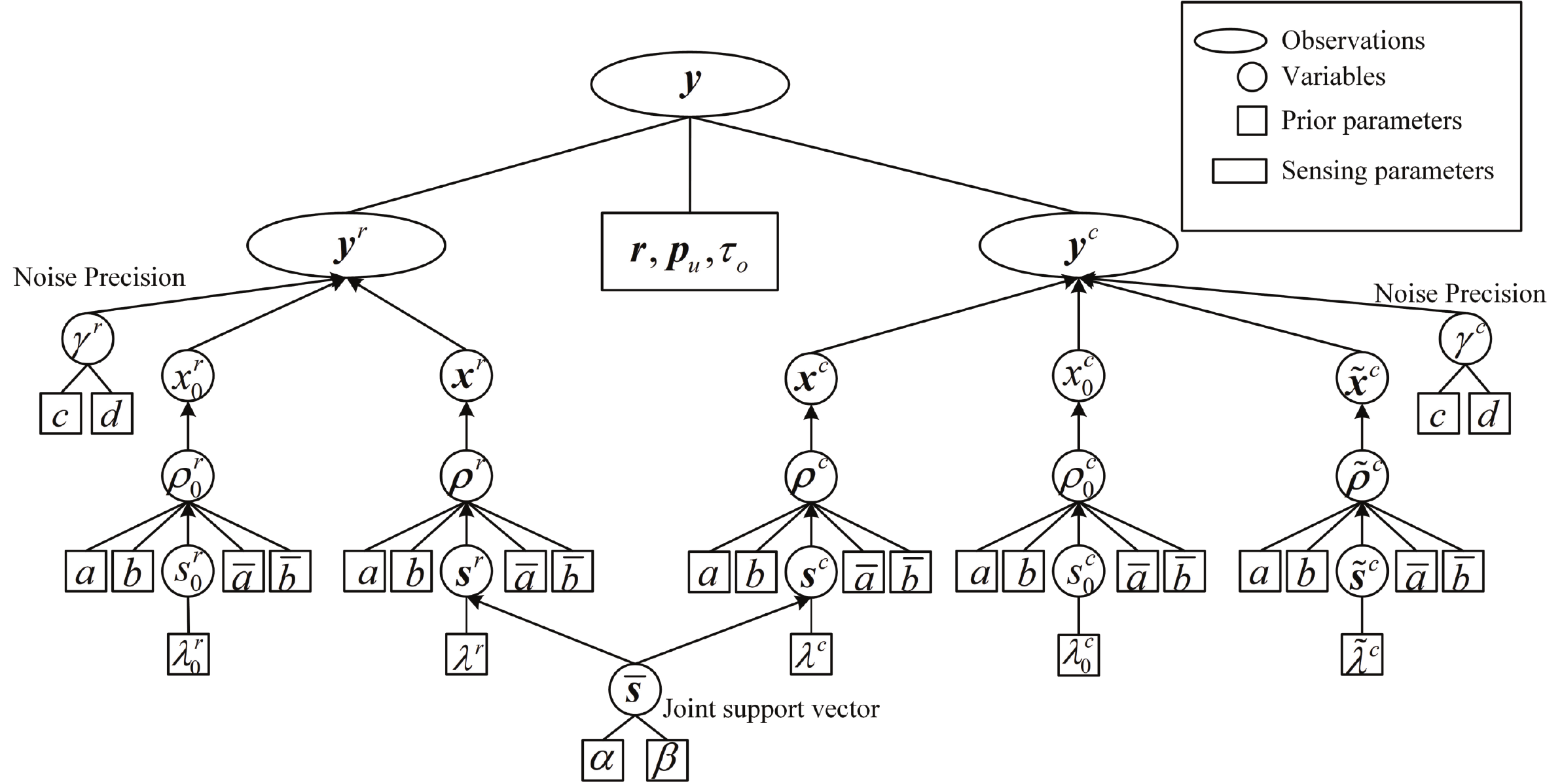}
\par\end{centering}
\caption{\label{fig:sparse_prior_model}Illustration of the three-layer sparse
prior model.}
\end{figure*}

\subsection{Location Domain Sparse Representation of Channels}

We introduce a grid-based solution to obtain a sparse representation
of the channels for better sensing and estimation performance. Specifically,
we first define a 2-D uniform grid $\left\{ \overline{\boldsymbol{r}}_{1},\ldots,\overline{\boldsymbol{r}}_{Q}\right\} \subset\mathcal{R}$
with size $H\times W$ of $Q\gg K+L$ positions for localizing the
radar targets and communication scatterers, as illustrated in Fig.
\ref{fig:Illustration-of-radar}, where the position grid points are
in a square area with $H$ rows and $W$ columns. Then, we define
a fixed grid $\left\{ \overline{\theta}_{1},...,\overline{\theta}_{U}\right\} $
of $U$ AoA points and a uniform grid $\left\{ \overline{\tau}_{1},...,\overline{\tau}_{V}\right\} \subset\left[\tau_{min},\tau_{max}\right]$
of $V$ time-of-arrival (ToA) points to estimate the multiple-bounce
NLoS channel vector, where $\left\{ \sin\overline{\theta}_{u}\right\} _{u=1}^{U}$
are uniformly distributed in the range $\left[-1,1\right]$ and the
delay plus time offsets of multiple-bounce NLoS paths, $\tilde{\tau}_{j}^{c}+\tau_{o},\forall j$,
are assumed to be within the range $\left[\tau_{min},\tau_{max}\right]$.

In practice, the true positions/AoAs/ToAs usually do not lie exactly
on the $Q$/$U$/$V$ discrete position/angle/delay grid points. In
this case, there will be an energy leakage effect, and thus we cannot
obtain an exact sparse representation of the corresponding channels.
The total energy of multiple-bounce NLoS paths is usually small compared
to the total energy of LoS and single-bounce NLoS paths. Therefore,
the energy leakage effect caused by the AoA and ToA mismatches is
negligible compared to the noise power. However, it is essential to
overcome the position mismatches for high-resolution localization.
One common solution is to introduce a dynamic position grid, denoted
by $\boldsymbol{r}\triangleq\left[\boldsymbol{r}_{1};\ldots;\boldsymbol{r}_{Q}\right]$,
instead of only using a fixed position grid. In this case, there always
exists an $\boldsymbol{r}^{*}$ that covers the true position of all
targets and communication scatterers. In general, the uniform grid
is chosen as the initial point for $\boldsymbol{r}$ in the algorithm,
which makes it easier to find a good solution for the non-convex MAP
estimation problem \cite{Huangzhe_JRC2}.

Then we define the sparse basis with a dynamic position grid for the
radar channel matrix and the single-bounce NLoS communication channel
vector as
\begin{equation}
\mathbf{A}\left(\boldsymbol{r}\right)\triangleq\left[\boldsymbol{a}\left(\theta^{r}\left(\boldsymbol{r}_{1}\right)\right),\ldots,\boldsymbol{a}\left(\theta^{r}\left(\boldsymbol{r}_{Q}\right)\right)\right]\in\mathbb{C}^{M\times Q}.
\end{equation}
Based on the angular and delay domain grids, we define the on-grid
basis for the multiple-bounce NLoS communication channel vector as
\begin{equation}
\begin{aligned}\overline{\mathbf{A}} & \triangleq\left[\boldsymbol{a}\left(\overline{\theta}_{1}\right),\ldots,\boldsymbol{a}\left(\overline{\theta}_{U}\right)\right]\in\mathbb{C}^{M\times U},\\
\overline{\mathbf{D}} & \triangleq\left[\boldsymbol{d}\left(\overline{\tau}_{1}\right),\ldots,\boldsymbol{d}\left(\overline{\tau}_{V}\right)\right]\in\mathbb{C}^{\left|\mathcal{N}_{u}\right|\times V},
\end{aligned}
\end{equation}
where $\boldsymbol{d}\left(\tau\right)\triangleq\left[e^{-j2\pi nf_{0}\tau}\right]_{n\in\mathcal{N}_{u}}\in\mathbb{C}^{\left|\mathcal{N}_{u}\right|\times1}$
denotes the delay domain basis vector.

The sparse representation of the radar channel matrix and the single-bounce
NLoS communication channel vector on the $n\textrm{-th}$ subcarrier
corresponding to (\ref{eq:Hnr}) and (\ref{eq:hncNL}) are respectively
given by
\begin{align}
\mathbf{H}_{n}^{r} & =\mathbf{A}\left(\boldsymbol{r}\right)\mathbf{D}_{n}^{r}\textrm{diag}\left(\boldsymbol{x}^{r}\right)\mathbf{A}^{T}\left(\boldsymbol{r}\right)\nonumber \\
 & +x_{0}^{r}e^{-j2\pi nf_{0}\left(\tau^{r}\left(\boldsymbol{p}_{u}\right)\right)}\boldsymbol{a}\left(\theta^{r}\left(\boldsymbol{p}_{u}\right)\right)\boldsymbol{a}^{T}\left(\theta^{r}\left(\boldsymbol{p}_{u}\right)\right),\label{eq:Hnr2}\\
\boldsymbol{h}_{n}^{c,\textrm{SL}} & =\mathbf{A}\left(\boldsymbol{r}\right)\mathbf{D}_{n}^{c}\boldsymbol{x}^{c},\label{eq:hncNL2}
\end{align}
where $\mathbf{D}_{n}^{r}$ and $\mathbf{D}_{n}^{c}$ are diagonal
matrices with the $q\textrm{-th}$ diagonal elements being $e^{-j2\pi nf_{0}\tau^{r}\left(\boldsymbol{r}_{q}\right)}$
and $e^{-j2\pi nf_{0}\left(\tau^{c}\left(\boldsymbol{r}_{q},\boldsymbol{p}_{u}\right)+\tau_{o}\right)}$,
respectively, $\boldsymbol{x}^{r}\in\mathbb{C}^{Q\times1}$ and $\boldsymbol{x}^{c}\in\mathbb{C}^{Q\times1}$
are called the location domain sparse radar channel vector\footnote{Note that we treat the echo signal reflected by the user separately
in (\ref{eq:Hnr2}) because the BS often has more accurate prior information
about the user location and the communication channel also depends
on the user location in a very different way compared to the position
of communication scatterers.} and single-bounce NLoS communication channel vector. $\boldsymbol{x}^{r}$
and $\boldsymbol{x}^{c}$ only have a few non-zero elements corresponding
to the position of targets and communication scatterers, respectively.
Specifically, the $q\textrm{-th}$ element of $\boldsymbol{x}^{r}$,
denoted by $x_{q}^{r}$, represents the complex reflection coefficient
of a target lying in the position $\boldsymbol{r}_{q}$. The $q\textrm{-th}$
element of $\boldsymbol{x}^{c}$, denoted by $x_{q}^{c}$, represents
the complex channel gain of the channel path with the corresponding
communication scatterer lying in the position $\boldsymbol{r}_{q}$.

Using the on-gird basis $\overline{\mathbf{A}}$ and $\overline{\mathbf{D}}$,
the multiple-bounce NLoS communication channel vectors on all subcarriers
can be expressed as
\begin{align}
\left[\boldsymbol{h}_{n}^{c,\textrm{ML}}\right]_{n\in\mathcal{N}_{u}}= & \textrm{vec}\left(\overline{\mathbf{A}}\widetilde{\mathbf{X}}^{c}\overline{\mathbf{D}}^{T}\right)=\left(\overline{\mathbf{D}}\otimes\overline{\mathbf{A}}\right)\widetilde{\boldsymbol{x}}^{c},\label{eq:hcMNL2}
\end{align}
where $\widetilde{\mathbf{X}}^{c}\in\mathbb{C}^{U\times V}$ is the
delay-angular domain sparse multiple-bounce NLoS communication channel
matrix, and $\widetilde{\boldsymbol{x}}^{c}\triangleq\textrm{vec}\left(\widetilde{\mathbf{X}}^{c}\right)\in\mathbb{C}^{UV\times1}$.

\subsection{Markov Random Field for 2-D Joint Burst Sparsity}

We shall introduce a three-layer sparse prior model, where a Markov
random field model is used to capture the 2-D joint burst sparsity
of the location domain channels. Specifically, let $\boldsymbol{\rho}^{r}\triangleq\left[\rho_{1}^{r},\ldots,\rho_{Q}^{r}\right]^{T}$
and $\boldsymbol{\rho}^{c}\triangleq\left[\rho_{1}^{c},\ldots,\rho_{Q}^{c}\right]^{T}$
represent the precision vectors of $\boldsymbol{x}^{r}$ and $\boldsymbol{x}^{c}$,
respectively, where $1/\rho_{q}^{r}$ and $1/\rho_{q}^{c}$ are the
variance of $x_{q}^{r}$ and $x_{q}^{c}$, respectively. Let $\boldsymbol{s}^{r}\triangleq\left[s_{1}^{r},\ldots,s_{Q}^{r}\right]^{T}\in\left\{ -1,1\right\} ^{Q}$
and $\boldsymbol{s}^{c}\triangleq\left[s_{1}^{c},\ldots,s_{Q}^{c}\right]^{T}\in\left\{ -1,1\right\} ^{Q}$
represent the support vectors of $\boldsymbol{x}^{r}$ and $\boldsymbol{x}^{c}$,
respectively. If there is a radar target (communication scatterer)
around the $q\textrm{-th}$ position grid $\boldsymbol{r}_{q}$, we
have $s_{q}^{r}=1$ and $x_{q}^{r}$ is non-zero ($s_{q}^{c}=1$ and
$x_{q}^{c}$ is non-zero). Otherwise, we have $s_{q}^{r}=-1$ and
$x_{q}^{r}=0$ ($s_{q}^{c}=-1$ and $x_{q}^{c}=0$). Then, we introduce
a joint support vector $\overline{\boldsymbol{s}}\triangleq\left[\overline{s}_{1},\ldots,\overline{s}_{Q}\right]^{T}\in\left\{ -1,1\right\} ^{Q}$
to represent the union of the positions of the radar targets and communication
scatterers. If either $s_{q}^{r}=1$ or $s_{q}^{c}=1$, we have $\overline{s}_{q}=1$.
Otherwise, we have $\overline{s}_{q}=-1$. The joint distribution
of $\boldsymbol{x}^{r}$, $\boldsymbol{x}^{c}$, $\boldsymbol{\rho}^{r}$,
$\boldsymbol{\rho}^{c}$, $\boldsymbol{s}^{r}$, $\boldsymbol{s}^{c}$,
and $\overline{\boldsymbol{s}}$ is represented as
\begin{align}
 & p\left(\boldsymbol{x}^{r},\boldsymbol{x}^{c},\boldsymbol{\rho}^{r},\boldsymbol{\rho}^{c},\boldsymbol{s}^{r},\boldsymbol{s}^{c},\overline{\boldsymbol{s}}\right)\\
= & \underbrace{p\left(\boldsymbol{s}^{r},\boldsymbol{s}^{c},\overline{\boldsymbol{s}}\right)}_{\textrm{Support}}\underbrace{p\left(\boldsymbol{\rho}^{r}\mid\boldsymbol{s}^{r}\right)p\left(\boldsymbol{\rho}^{c}\mid\boldsymbol{s}^{c}\right)}_{\textrm{Precision}}\underbrace{p\left(\boldsymbol{x}^{r}\mid\boldsymbol{\rho}^{r}\right)p\left(\boldsymbol{x}^{c}\mid\boldsymbol{\rho}^{c}\right)}_{\textrm{Sparse\ signal}}.\nonumber 
\end{align}
The three-layer sparse prior model is shown in Fig. \ref{fig:sparse_prior_model}.
A similar three-layer model has been considered in \cite{LiuAn_directloc_vehicles,LiuAn_CE_Turbo_VBI}
and is shown to be more flexible to capture the structured sparsity
of realistic channels.

The sparse signals $\boldsymbol{x}^{r}$ and $\boldsymbol{x}^{c}$
follow complex Gaussian distributions with zero mean and variance
$1/\boldsymbol{\rho}^{r}$ and $1/\boldsymbol{\rho}^{c}$, respectively.
Moreover, conditioned on $\boldsymbol{\rho}^{r}$ and $\boldsymbol{\rho}^{c}$,
the elements of $\boldsymbol{x}^{r}$ and $\boldsymbol{x}^{c}$ are
assumed to be independent, i.e.,
\begin{equation}
\begin{aligned}p\left(\boldsymbol{x}^{r}\mid\boldsymbol{\rho}^{r}\right) & =\prod_{q}p\left(x_{q}^{r}\mid\rho_{q}^{r}\right)=\prod_{q}\mathcal{CN}\left(x_{q}^{r};0,1/\rho_{q}^{r}\right),\\
p\left(\boldsymbol{x}^{c}\mid\boldsymbol{\rho}^{c}\right) & =\prod_{q}p\left(x_{q}^{c}\mid\rho_{q}^{c}\right)=\prod_{q}\mathcal{CN}\left(x_{q}^{c};0,1/\rho_{q}^{c}\right).
\end{aligned}
\end{equation}
The conditional distributions $p\left(\boldsymbol{\rho}^{r}\mid\boldsymbol{s}^{r}\right)$
and $p\left(\boldsymbol{\rho}^{c}\mid\boldsymbol{s}^{c}\right)$ are
respectively given by
\begin{equation}
\begin{aligned}p\left(\boldsymbol{\rho}^{r}\mid\boldsymbol{s}^{r}\right)= & \prod_{q}\left(\delta\left(s_{q}^{r}-1\right)\textrm{Gamma}\left(\rho_{q}^{r};a,b\right)\right.\\
+ & \left.\delta\left(s_{q}^{r}+1\right)\textrm{Gamma}\left(\rho_{q}^{r};\overline{a},\overline{b}\right)\right),\\
p\left(\boldsymbol{\rho}^{c}\mid\boldsymbol{s}^{c}\right)= & \prod_{q}\left(\delta\left(s_{q}^{c}-1\right)\textrm{Gamma}\left(\rho_{q}^{c};a,b\right)\right.\\
+ & \left.\delta\left(s_{q}^{c}+1\right)\textrm{Gamma}\left(\rho_{q}^{c};\overline{a},\overline{b}\right)\right),
\end{aligned}
\end{equation}
where $\delta\left(\cdot\right)$ is the Dirac Delta function. When
$s_{q}^{r}=1$, $x_{q}^{r}$ is a non-zero element and the corresponding
variance $1/\rho_{q}^{r}$ is $\Theta\left(1\right)$. In this case,
$a$ and $b$ should be chosen to satisfy $\tfrac{a}{b}=\mathbb{E}\left(\rho_{q}^{r}\right)=\Theta\left(1\right)$.
When $s_{q}^{r}=-1$, $x_{q}^{r}$ is a zero element and the corresponding
variance $1/\rho_{q}^{r}$ is close to zero. In this case, $\overline{a}$
and $\overline{b}$ should be chosen to satisfy $\tfrac{\overline{a}}{\overline{b}}=\mathbb{E}\left(\rho_{q}^{r}\right)\gg1$.
A typical value is $a=1$, $b=1$, $\overline{a}=1$, and $\overline{b}=10^{-5}$
\cite{LiuAn_CE_Turbo_VBI}. Since the gamma prior is conjugate to
the Gaussian prior, we can derive the close-form expressions when
performing Bayesian inference. The details will be elaborated in Section
\ref{sec:SEA-Turbo-SBI-Algorithm}.

The joint distribution of support vectors can be further decomposed
into
\begin{equation}
\begin{aligned}p\left(\boldsymbol{s}^{r},\boldsymbol{s}^{c},\overline{\boldsymbol{s}}\right) & =p\left(\boldsymbol{s}^{r}\mid\overline{\boldsymbol{s}}\right)p\left(\boldsymbol{s}^{c}\mid\overline{\boldsymbol{s}}\right)p\left(\overline{\boldsymbol{s}}\right)\\
 & =\prod_{q}p\left(s_{q}^{r}\mid\overline{s}_{q}\right)\prod_{q}p\left(s_{q}^{c}\mid\overline{s}_{q}\right)p\left(\overline{\boldsymbol{s}}\right),
\end{aligned}
\end{equation}
where the conditional distributions are given by
\begin{align}
p\left(s_{q}^{r}\mid\overline{s}_{q}\right) & =\delta\left(\overline{s}_{q}+1\right)\delta\left(s_{q}^{r}+1\right)\\
 & +\delta\left(\overline{s}_{q}-1\right)\left(\delta\left(s_{q}^{r}-1\right)\lambda{}_{q}^{r}+\delta\left(s_{q}^{r}+1\right)\left(1-\lambda{}_{q}^{r}\right)\right),\nonumber \\
p\left(s_{q}^{c}\mid\overline{s}_{q}\right) & =\delta\left(\overline{s}_{q}+1\right)\delta\left(s_{q}^{c}+1\right)\nonumber \\
 & +\delta\left(\overline{s}_{q}-1\right)\left(\delta\left(s_{q}^{c}-1\right)\lambda_{q}^{c}+\delta(s_{q}^{c}+1)\left(1-\lambda_{q}^{c}\right)\right),\nonumber 
\end{align}
where $\lambda{}_{q}^{r}$ and $\lambda_{q}^{c}$ represent the probability
of $s_{q}^{r}=1$ and $s_{q}^{c}=1$ conditioned on $\overline{s}_{q}=1$,
respectively.

Moreover, we use a spatially non-stationary Markov random field model
to describe the 2-D joint burst sparsity of $\boldsymbol{x}^{r}$
and $\boldsymbol{x}^{c}$. Based on the Ising model \cite{MRF1,MRF2},
the joint support vector can be modeled as
\begin{equation}
\begin{aligned}p\left(\overline{\boldsymbol{s}}\right) & =\frac{1}{Z\left(\boldsymbol{\zeta}\right)}\exp\left(\sum_{q=1}^{Q}\left(\frac{1}{2}\sum_{i\in\mathcal{N}_{q}}\beta_{iq}\overline{s}_{i}-\alpha_{q}\right)\overline{s}_{q}\right)\\
 & =\frac{1}{Z\left(\boldsymbol{\zeta}\right)}\left(\stackrel[q=1]{Q}{\prod}\underset{i\in\mathcal{N}_{q}}{\prod}\varphi\left(\overline{s}_{q},\overline{s}_{i}\right)\right)^{\frac{1}{2}}\prod_{q=1}^{Q}\psi\left(\overline{s}_{q}\right),
\end{aligned}
\label{eq:p(s)}
\end{equation}
where $\varphi\left(\overline{s}_{q},\overline{s}_{i}\right)=\exp\left(\beta_{iq}\overline{s}_{i}\overline{s}_{q}\right)$,
$\psi\left(\overline{s}_{q}\right)=\exp\left(-\alpha_{q}\overline{s}_{q}\right)$,
$\mathcal{N}_{q}$ denotes the index set for the neighbors of $\overline{s}_{q}$,
$\boldsymbol{\zeta}\triangleq\left\{ \alpha_{q},\beta_{iq}\mid i\in\mathcal{N}_{q},\forall q\right\} $
denotes the MRF parameters, and $Z\left(\boldsymbol{\zeta}\right)$
denotes the partition function. Since $\alpha_{q}$ and $\beta_{iq}$
depends on the position index $q$, the MRF in (\ref{eq:p(s)}) is
spatially non-stationary, which helps to model scattering clusters
with diverse random sizes and positions. Specifically, a higher value
of $\beta_{iq}$ implies a larger size for non-zero bursts, and a
higher value of $\alpha_{q}$ implies sparser signal activity.
\begin{figure}[t]
\begin{centering}
\includegraphics[clip,width=80mm]{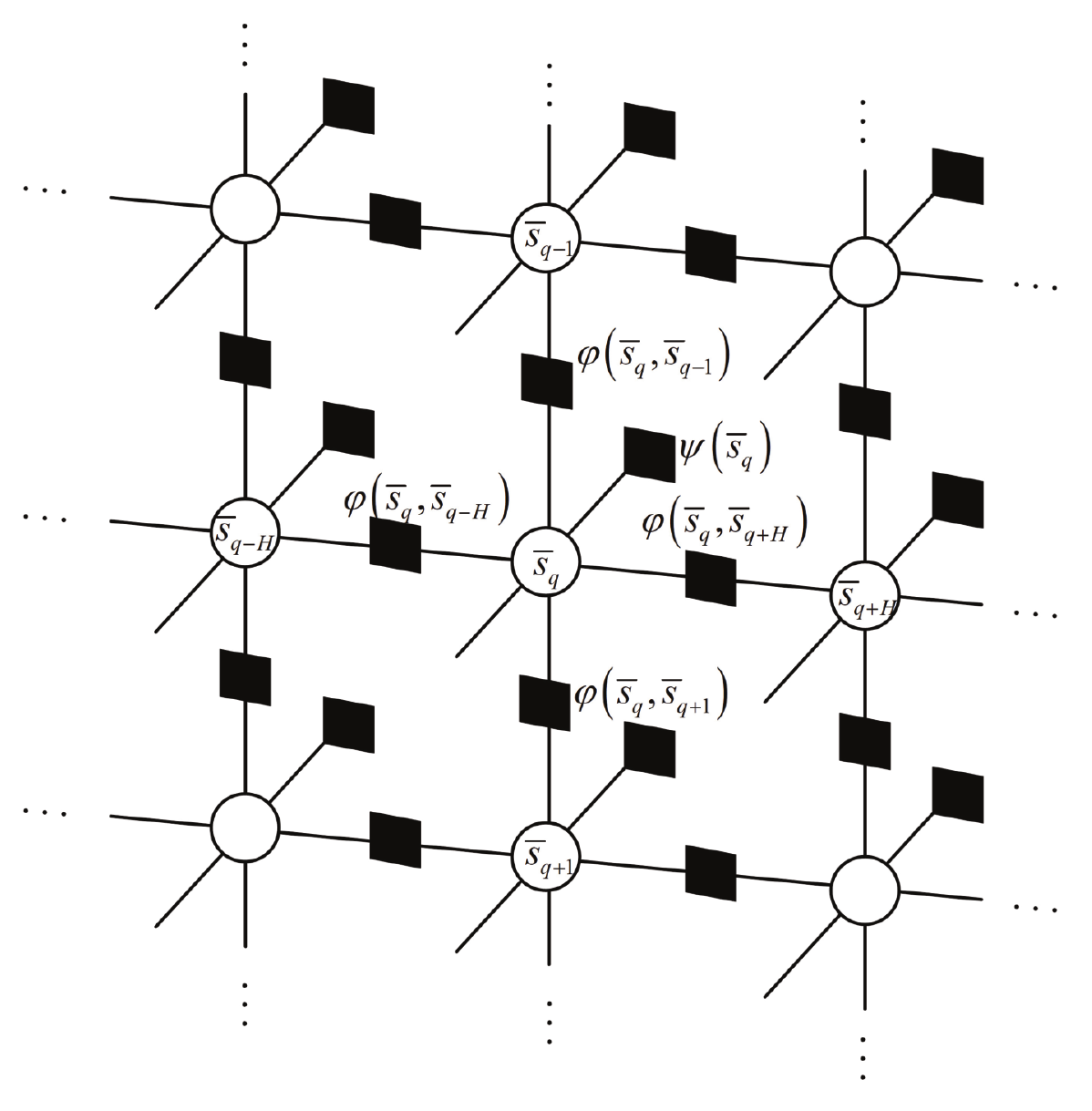}\caption{\label{fig:Factor-graph-of-MRF}Factor graph of the 4-connected MRF
model.}
\par\end{centering}
\end{figure}

We try to recover the joint support vector $\overline{\boldsymbol{s}}$
associated with the 2-D position grid of $Q=H\times W$ points. In
this case, the 4-connected MRF model is suitable to process such a
2-D sparse signal recovery problem. The factor graph of the 4-connected
MRF model is shown in Fig. \ref{fig:Factor-graph-of-MRF}. In the
MRF model, the variable nodes $\left\{ \overline{s}_{q}\right\} _{q=1}^{Q}$
are scheduled in $H$ rows and $W$ columns. Most of the variable
nodes have four neighboring nodes, except for the nodes at the boundaries.
To be specific, the left, right, top, and bottom neighboring nodes
of $\overline{s}_{q}$ are $\overline{s}_{q-H}$, $\overline{s}_{q+H}$,
$\overline{s}_{q-1}$, and $\overline{s}_{q+1}$, respectively. Two
types of factor nodes are involved in the factor graph. The factor
node $\varphi\left(\overline{s}_{q},\overline{s}_{i}\right)$ connecting
$\overline{s}_{q}$ and $\overline{s}_{i}$ describes the correlation
between two neighboring nodes, while the factor node $\psi\left(\overline{s}_{q}\right)$
connected to $\overline{s}_{q}$ directly affects the sparse probability
of $\overline{s}_{q}$.

To automatically learn the noise precision, we assume a gamma distribution
with shape parameter $c$ and rate parameter $d$ as the prior for
$\gamma^{r}$ and $\gamma^{c}$, i.e.,
\begin{equation}
\begin{aligned}p\left(\gamma^{r}\right) & =\textrm{Gamma}\left(\gamma^{r};c,d\right),\\
p\left(\gamma^{c}\right) & =\textrm{Gamma}\left(\gamma^{c};c,d\right).
\end{aligned}
\end{equation}

The sparse prior model for $x_{0}^{r}$, $x_{0}^{c}$, and $\widetilde{\boldsymbol{x}}^{c}$
is similar to that of $\boldsymbol{x}^{r}$ and $\boldsymbol{x}^{c}$
except that there is no joint support vector, as shown in Fig. \ref{fig:sparse_prior_model},
where $\rho_{0}^{r}$, $\rho_{0}^{c}$, and $\boldsymbol{\widetilde{\rho}}^{c}$
denote the precision of $x_{0}^{r}$, $x_{0}^{c}$, and $\widetilde{\boldsymbol{x}}^{c}$,
respectively, $s_{0}^{r},$ $s_{0}^{c}$, and $\widetilde{\boldsymbol{s}}^{c}$
denote the support of $x_{0}^{r}$, $x_{0}^{c}$, and $\widetilde{\boldsymbol{x}}^{c}$,
respectively, $\lambda_{0}^{r}$, $\lambda_{0}^{r}$, and $\widetilde{\lambda_{j}}^{c},\forall j$
give the probability of $s_{0}^{r}=1$, $s_{0}^{c}=1$, and $\widetilde{s}_{j}^{c}=1,\forall j$,
respectively.

The unknown parameters of the probability model, denoted by $\left\{ \lambda{}_{q}^{r},\lambda{}_{q}^{c},\lambda_{0}^{r},\lambda_{0}^{c},\widetilde{\lambda_{j}}^{c},\boldsymbol{\zeta}\right\} $,
can be automatically learned based on the EM method. However, the
non-stationary MRF in (\ref{eq:p(s)}) is quite complicated and the
corresponding model parameters $\boldsymbol{\zeta}$ cannot be easily
learned via the conventional EM method. We will describe how to learn
MRF parameters approximately via a low-complexity method in Subsection
\ref{subsec:Turbo-IF-VBI-M-Step}.

\subsection{Sparse Bayesian Inference with Uncertain Parameters}

Using the location domain sparse representation in (\ref{eq:Hnr2})
and (\ref{eq:hncNL2}) and the angle-delay domain sparse representation
in (\ref{eq:hcMNL2}), the reflected downlink pilot signal and received
uplink pilot signal on all available subcarriers can be expressed
as
\begin{subequations}
\begin{align}
\boldsymbol{y}^{r} & =\boldsymbol{\Phi}^{r}\left(\boldsymbol{r},\boldsymbol{p}_{u}\right)\left[x_{0}^{r},\left(\boldsymbol{x}^{r}\right)^{T}\right]^{T}+\boldsymbol{z}^{r},\label{eq:yr}\\
\boldsymbol{y}^{c} & =\boldsymbol{\Phi}^{c}\left(\boldsymbol{r},\boldsymbol{p}_{u},\tau_{o}\right)\left[x_{0}^{c},\left(\boldsymbol{x}^{c}\right)^{T},\left(\widetilde{\boldsymbol{x}}^{c}\right)^{T}\right]^{T}+\boldsymbol{z}^{c},\label{eq:yc}
\end{align}
\end{subequations}
where $\boldsymbol{y}^{r}\triangleq\left[\boldsymbol{y}_{n}^{r}\right]_{n\in\mathcal{N}_{b}}\in\mathbb{C}^{M\left|\mathcal{N}_{b}\right|\times1}$,
$\boldsymbol{y}^{c}\triangleq\left[\boldsymbol{y}_{n}^{c}\right]_{n\in\mathcal{N}_{u}}\in\mathbb{C}^{M\left|\mathcal{N}_{u}\right|\times1}$,
$\boldsymbol{z}^{r}\triangleq\left[\boldsymbol{z}_{n}^{r}\right]_{n\in\mathcal{N}_{b}}\in\mathbb{C}^{M\left|\mathcal{N}_{b}\right|\times1}$,
and $\boldsymbol{z}^{c}\triangleq\left[\boldsymbol{z}_{n}^{c}\right]_{n\in\mathcal{N}_{u}}\in\mathbb{C}^{M\left|\mathcal{N}_{u}\right|\times1}$.

The radar and communication measurement matrices in (\ref{eq:yr})
and (\ref{eq:yc}) can be decomposed into some submatrices, respectively,
i.e.,
\begin{equation}
\begin{aligned}\boldsymbol{\Phi}^{r}\left(\boldsymbol{r},\boldsymbol{p}_{u}\right) & \triangleq\left[\boldsymbol{\Phi}^{r,0},\boldsymbol{\Phi}^{r,1}\right]\in\mathbb{C}^{M\left|\mathcal{N}_{b}\right|\times\left(Q+1\right)},\\
\boldsymbol{\Phi}^{c}\left(\boldsymbol{r},\boldsymbol{p}_{u},\tau_{o}\right) & \triangleq\left[\boldsymbol{\Phi}^{c,0},\boldsymbol{\Phi}^{c,1},\boldsymbol{\Phi}^{c,2}\right]\in\mathbb{C}^{M\left|\mathcal{N}_{u}\right|\times\left(Q+UV+1\right)},
\end{aligned}
\end{equation}
where
\begin{align}
\boldsymbol{\Phi}^{r,0} & =\left[e^{-j2\pi nf_{0}\left(\tau^{r}\left(\boldsymbol{p}_{u}\right)\right)}\boldsymbol{a}\left(\theta^{r}\left(\boldsymbol{p}_{u}\right)\right)\boldsymbol{a}^{T}\left(\theta^{r}\left(\boldsymbol{p}_{u}\right)\right)\boldsymbol{v}_{n}^{r}\right]_{n\in\left|\mathcal{N}_{b}\right|},\nonumber \\
\boldsymbol{\bar{\Phi}}^{r,1} & =\left[\left(\left(\boldsymbol{v}_{n}^{r}\right)^{T}\mathbf{A}\left(\boldsymbol{r}\right)\right)\otimes\left(\mathbf{A}\left(\boldsymbol{r}\right)\mathbf{D}_{n}^{r}\right)\right]_{n\in\left|\mathcal{N}_{b}\right|},\nonumber \\
\boldsymbol{\Phi}^{c,0} & =\left[u_{n}^{c}e^{-j2\pi nf_{0}\tau_{o}}\boldsymbol{a}\left(\theta^{c}\left(\boldsymbol{p}_{u}\right)\right)\right]_{n\in\left|\mathcal{N}_{u}\right|},\nonumber \\
\boldsymbol{\Phi}^{c,1} & =\left[u_{n}^{c}\mathbf{A}\left(\boldsymbol{r}\right)\mathbf{D}_{n}^{c}\right]_{n\in\left|\mathcal{N}_{u}\right|},\nonumber \\
\boldsymbol{\Phi}^{c,2} & =\left(\textrm{diag}\left(\left[u_{n}^{c}\right]_{n\in\left|\mathcal{N}_{u}\right|}\right)\mathbf{\overline{D}}\right)\otimes\mathbf{\overline{A}},
\end{align}
where $\boldsymbol{\Phi}^{r,1}\in\mathbb{C}^{M\left|\mathcal{N}_{b}\right|\times Q}$
consists of the $\left(\left(q-1\right)Q+q\right)\textrm{-th}$ column
of $\boldsymbol{\bar{\Phi}}^{r,1}$ for $q=1,\ldots,Q$.

For convenience, we combine (\ref{eq:yr}) and (\ref{eq:yc}) into
a linear observation model as
\begin{equation}
\boldsymbol{y}=\boldsymbol{\Phi}\left(\boldsymbol{\vartheta}\right)\boldsymbol{x}+\boldsymbol{z},\label{eq:y}
\end{equation}
where $\boldsymbol{\xi}\triangleq\left\{ \boldsymbol{r},\boldsymbol{p}_{u},\tau_{o}\right\} $
is the collection of sensing parameters, $\boldsymbol{y}\triangleq\left[\left(\boldsymbol{y}^{r}\right)^{T},\left(\boldsymbol{y}^{c}\right)^{T}\right]^{T}$,
$\boldsymbol{z}\triangleq\left[\left(\boldsymbol{z}^{r}\right)^{T},\left(\boldsymbol{z}^{c}\right)^{T}\right]^{T}$,
$\boldsymbol{x}\triangleq\left[x_{0}^{r},\left(\boldsymbol{x}^{r}\right)^{T},x_{0}^{c},\left(\boldsymbol{x}^{c}\right)^{T},\left(\widetilde{\boldsymbol{x}}^{c}\right)^{T}\right]^{T}$,
and $\boldsymbol{\Phi}\left(\boldsymbol{\vartheta}\right)\triangleq\mathrm{\textrm{BlockDiag}}\left(\boldsymbol{\Phi}^{r}\left(\boldsymbol{\vartheta}\right),\boldsymbol{\Phi}^{c}\left(\boldsymbol{\vartheta}\right)\right)$.

To simply the notation, the precision vector and the support vector
of $\boldsymbol{x}$ are respectively defined as
\begin{align*}
\boldsymbol{\rho} & \triangleq\left[\rho_{0}^{r},\left(\boldsymbol{\rho}^{r}\right)^{T},\rho_{0}^{c},\left(\boldsymbol{\rho}^{c}\right)^{T},\left(\widetilde{\boldsymbol{\rho}}^{c}\right)^{T}\right]^{T},\\
\boldsymbol{s} & \triangleq\left[s_{0}^{r},\left(\boldsymbol{s}^{r}\right)^{T},s_{0}^{c},\left(\boldsymbol{s}^{c}\right)^{T},\left(\widetilde{\boldsymbol{s}}^{c}\right)^{T}\right]^{T}.
\end{align*}
Our primary goal is to estimate the channel vector $\boldsymbol{x}$,
the support vector $\boldsymbol{s}$, and the uncertain parameters
$\boldsymbol{\xi}\triangleq\left\{ \boldsymbol{\vartheta},\boldsymbol{\zeta}\right\} $
given observation $\boldsymbol{y}$ in model (\ref{eq:y}). To be
specific, for given $\boldsymbol{\xi}$, we aim at computing the conditional
marginal posteriors, i.e., $p\left(\boldsymbol{x}\mid\boldsymbol{y};\boldsymbol{\xi}\right)$
and $p\left(s_{i}\mid\boldsymbol{y};\boldsymbol{\xi}\right),\forall i$.
On the other hand, the uncertain parameters $\boldsymbol{\xi}$ are
obtained by the MAP estimator as follows:
\begin{equation}
\begin{aligned}\boldsymbol{\xi}^{\ast} & =\underset{\boldsymbol{\xi}}{\arg\max}\ln p\left(\boldsymbol{\xi}\mid\boldsymbol{y}\right)\\
 & =\underset{\boldsymbol{\xi}}{\arg\max}\sum_{\overline{\boldsymbol{s}}}\ln\int_{\boldsymbol{v}}p\left(\boldsymbol{y},\boldsymbol{v},\overline{\boldsymbol{s}};\boldsymbol{\xi}\right)p\left(\boldsymbol{\vartheta}\right),
\end{aligned}
\label{eq:M-step}
\end{equation}
where $\boldsymbol{v}\triangleq\left\{ \boldsymbol{x},\boldsymbol{\rho},\boldsymbol{s},\gamma^{r},\gamma^{c}\right\} $
and $p\left(\boldsymbol{\vartheta}\right)$ denotes the known prior
distribution of $\boldsymbol{\vartheta}$. Once we obtain the MAP
estimate of $\boldsymbol{\xi}^{\ast}$, we can obtain the minimum
mean square error (MMSE) estimate of $\boldsymbol{x}$ as $\boldsymbol{x}^{*}=\int_{\boldsymbol{x}}\boldsymbol{x}p\left(\boldsymbol{x}\mid\boldsymbol{y};\boldsymbol{\xi}^{\ast}\right)$
and the MAP estimate of $\boldsymbol{s}$ as $s_{i}^{\ast}=\arg\max_{s_{i}}p\left(s_{i}\mid\boldsymbol{y};\boldsymbol{\xi}^{\ast}\right),\forall i$.

However, the corresponding factor graph of the probability model contains
loops and the associated sparse Bayesian inference problem is NP-hard.
Therefore, it is exceedingly challenging to calculate the above conditional
marginal posteriors precisely. In the following section, we present
the Turbo-IF-VBI algorithm, which uses the turbo approach to calculate
approximate marginal posteriors and applies a variation of the EM
method to find an approximate solution for (\ref{eq:M-step}).

\section{Turbo-IF-VBI Algorithm\label{sec:SEA-Turbo-SBI-Algorithm}}

\subsection{Outline of the Turbo-IF-VBI Algorithm}

\begin{figure}[t]
\begin{centering}
\textsf{\includegraphics[width=1\columnwidth]{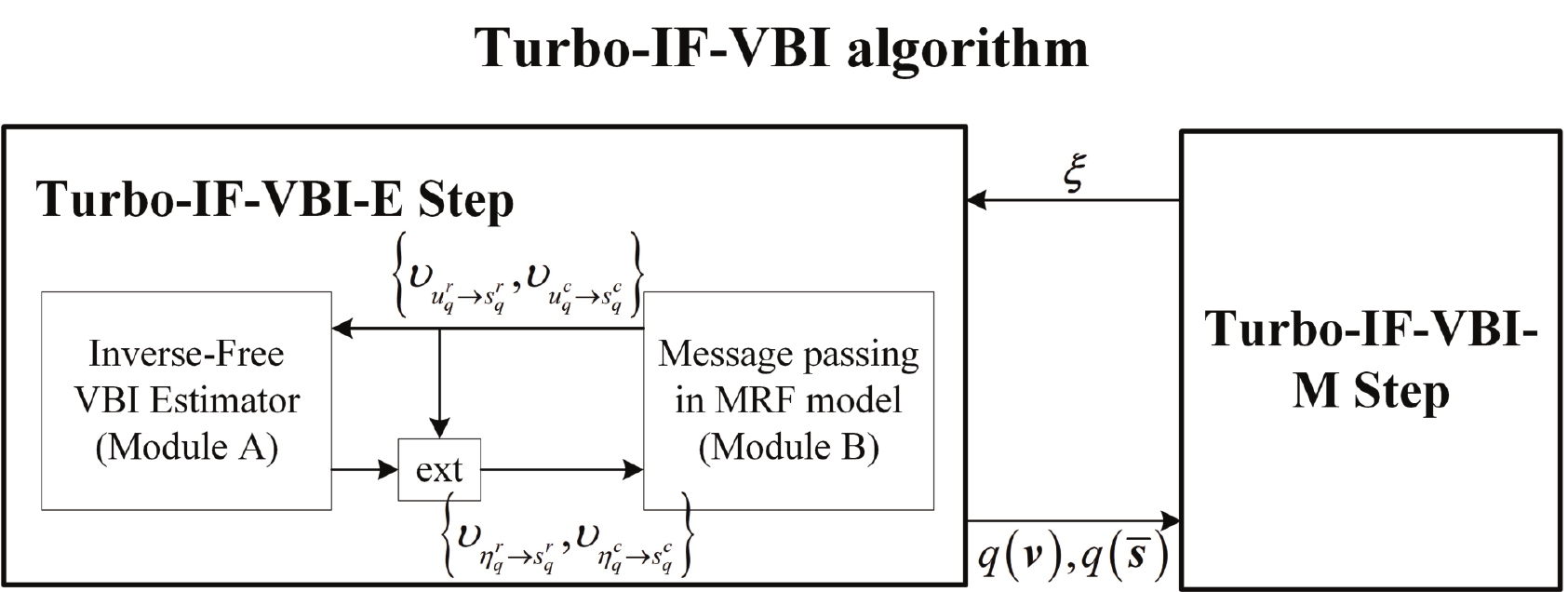}}
\par\end{centering}
\caption{\label{fig:Turbo-IF-VBI}Framework of the Turbo-IF-VBI algorithm.}
\end{figure}

The primary goal of the Turbo-IF-VBI algorithm is to simultaneously
maximize the marginal log-posterior $\ln p\left(\boldsymbol{y},\boldsymbol{\xi}\right)$
with respect to the uncertain parameters $\boldsymbol{\xi}$ in (\ref{eq:M-step})
and approximately calculate the conditional posteriors. As illustrated
in Fig. \ref{fig:Turbo-IF-VBI}, the Turbo-IF-VBI algorithm iterates
between the next two major steps until convergence.

\begin{itemize}
\item \textbf{Turbo-IF-VBI-E Step:} For given $\boldsymbol{\xi}^{\left(t\right)}$
in the $t\textrm{-th}$ iteration, calculate the approximate marginal
posteriors, denoted by $q\left(\boldsymbol{v}\mid\boldsymbol{y};\boldsymbol{\xi}^{\left(t\right)}\right)$
and $q\left(\overline{\boldsymbol{s}}\mid\boldsymbol{y};\boldsymbol{\xi}^{\left(t\right)}\right)$,
based on the turbo approach.
\item \textbf{Turbo-IF-VBI-M Step:} Construct a surrogate function for $\ln p\left(\boldsymbol{y},\boldsymbol{\xi}\right)$
based on $q\left(\boldsymbol{v}\mid\boldsymbol{y};\boldsymbol{\xi}^{\left(t\right)}\right)$
and $q\left(\overline{\boldsymbol{s}}\mid\boldsymbol{y};\boldsymbol{\xi}^{\left(t\right)}\right)$
obtained in the Turbo-IF-VBI-E Step, then maximize the surrogate function
with respect to $\boldsymbol{\xi}$.
\end{itemize}
The Turbo-IF-VBI-E Step is an inverse-free algorithm by combining
the IF-VBI estimator and message passing via the turbo framework,
where the IF-VBI avoids the matrix inverse operation via maximizing
a relaxed ELBO. Furthermore, the Turbo-IF-VBI-M Step is challenging
as the surrogate function constructed by the conventional EM method
involves exponential computational complexity. To overcome this challenge,
we propose a low-complexity method based on pseudo-likelihood approximation
to learn MRF parameters. In the following, we first elaborate on how
to approximately calculate $q\left(\boldsymbol{v}\mid\boldsymbol{y};\boldsymbol{\xi}^{\left(t\right)}\right)$
and $q\left(\overline{\boldsymbol{s}}\mid\boldsymbol{y};\boldsymbol{\xi}^{\left(t\right)}\right)$
in the Turbo-IF-VBI-E Step. Then, we show how to update $\boldsymbol{\xi}$
in the Turbo-IF-VBI-M Step.
\begin{figure}[t]
\centering{}\includegraphics[clip,width=1\columnwidth]{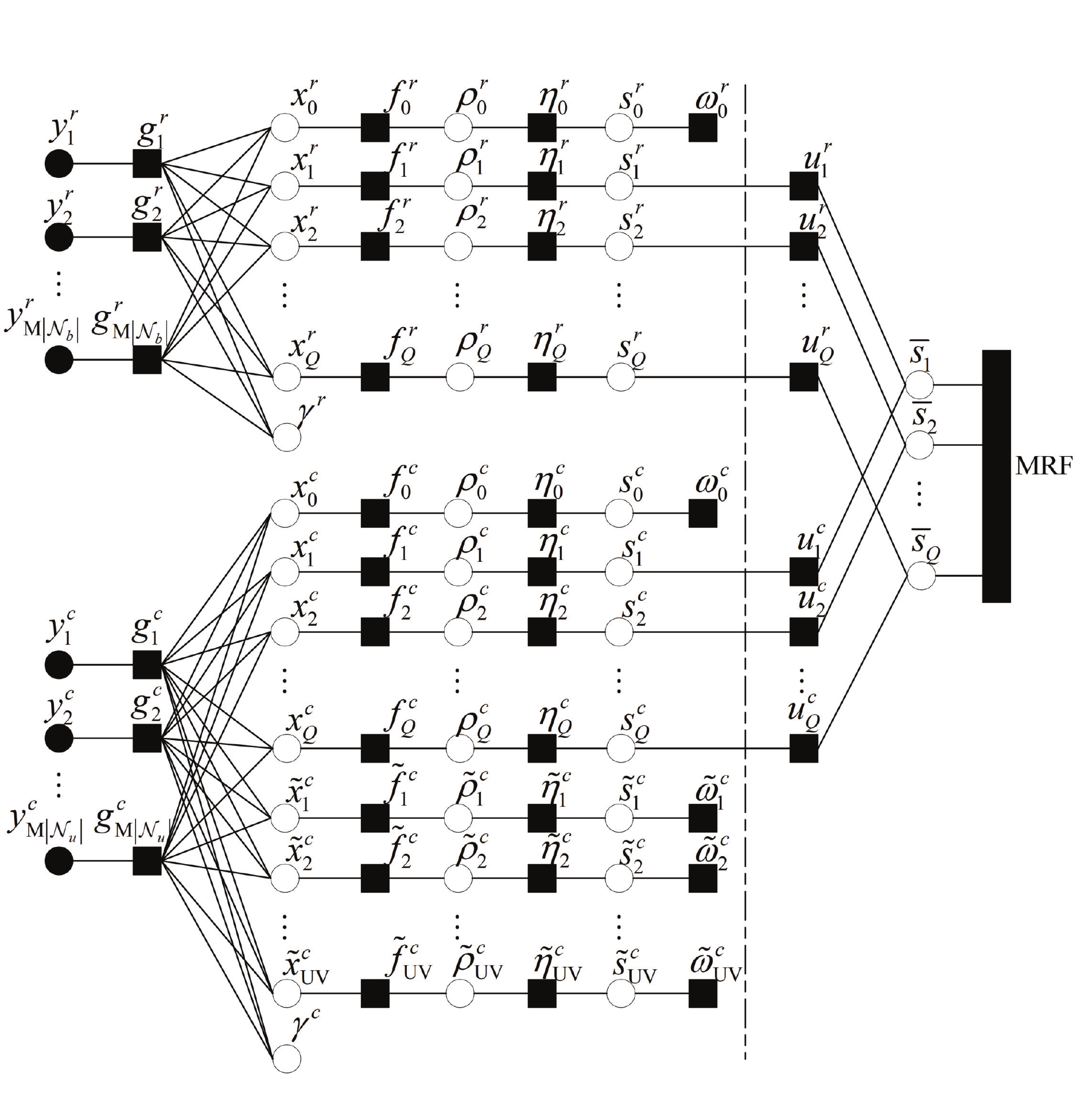}\caption{\label{fig:factor graph}Factor graph of the joint distribution $p\left(\boldsymbol{y},\boldsymbol{v},\overline{\boldsymbol{s}};\boldsymbol{\xi}\right)$.}
\end{figure}
\begin{table*}[t]
\caption{\label{tab:Factors_Table}Factors, Distributions and Functional forms
in Fig. \ref{fig:factor graph}. $\boldsymbol{\Phi}_{m}^{r}\left(\boldsymbol{\vartheta}\right)$
and $\boldsymbol{\Phi}_{m}^{c}\left(\boldsymbol{\vartheta}\right)$
denote the $m\textrm{-th}$ row of $\boldsymbol{\Phi}^{r}\left(\boldsymbol{\vartheta}\right)$
and $\boldsymbol{\Phi}^{c}\left(\boldsymbol{\vartheta}\right)$, respectively.}

\centering{}%
\begin{tabular}{|c|c|c|}
\hline 
Factor & Distribution & Functional form\tabularnewline
\hline 
\hline 
$\begin{array}{c}
g_{m}^{r}\\
g_{m}^{c}
\end{array}$ & $\begin{array}{c}
p\left(y_{m}^{r}\mid x_{0}^{r},\boldsymbol{x}^{r},\gamma^{r};\boldsymbol{\xi}\right)\\
p\left(y_{m}^{c}\mid x_{0}^{c},\boldsymbol{x}^{c},\widetilde{\boldsymbol{x}}^{c},\gamma^{c};\boldsymbol{\xi}\right)
\end{array}$ & $\begin{array}{c}
\mathcal{CN}\left(y_{m}^{r};\boldsymbol{\Phi}_{m}^{r}\left(\boldsymbol{\vartheta}\right)\left[x_{0}^{r},\left(\boldsymbol{x}^{r}\right)^{T}\right]^{T},1/\gamma^{r}\right)\\
\mathcal{CN}\left(y_{m}^{c};\boldsymbol{\Phi}_{m}^{c}\left(\boldsymbol{\vartheta}\right)\left[x_{0}^{c},\left(\boldsymbol{x}^{c}\right)^{T},\left(\widetilde{\boldsymbol{x}}^{c}\right)^{T}\right]^{T},1/\gamma^{c}\right)
\end{array}$\tabularnewline
\hline 
$\begin{array}{c}
f_{q}^{r}\\
f_{q}^{c}\\
\widetilde{f}_{j}^{c}
\end{array}$ & $\begin{array}{c}
p\left(x_{q}^{r}\mid\rho_{q}^{r}\right)\\
p\left(x_{q}^{c}\mid\rho_{q}^{c}\right)\\
p\left(x_{j}^{c}\mid\widetilde{\rho}_{j}^{c}\right)
\end{array}$ & $\begin{array}{c}
\mathcal{CN}\left(x_{q}^{r};0,1/\rho_{q}^{r}\right)\\
\mathcal{CN}\left(x_{q}^{c};0,1/\rho_{q}^{c}\right)\\
\mathcal{CN}\left(\widetilde{x}_{j}^{r};0,1/\widetilde{\rho}_{j}^{r}\right)
\end{array}$\tabularnewline
\hline 
$\begin{array}{c}
\eta_{q}^{r}\\
\eta_{q}^{c}\\
\widetilde{\eta}_{j}^{c}
\end{array}$ & $\begin{array}{c}
p\left(\rho_{q}^{r}\mid s_{q}^{r}\right)\\
p\left(\rho_{q}^{c}\mid s_{q}^{c}\right)\\
p\left(\tilde{\rho}_{j}^{c}\mid\widetilde{s}_{j}^{c}\right)
\end{array}$ & $\begin{array}{c}
\delta\left(s_{q}^{r}-1\right)\textrm{Gamma}\left(\rho_{q}^{r};a,b\right)+\delta\left(s_{q}^{r}+1\right)\textrm{Gamma}\left(\rho_{q}^{r};\overline{a},\overline{b}\right)\\
\delta\left(s_{q}^{c}-1\right)\textrm{Gamma}\left(\rho_{q}^{c};a,b\right)+\delta\left(s_{q}^{c}+1\right)\textrm{Gamma}\left(\rho_{q}^{c};\overline{a},\overline{b}\right)\\
\delta\left(\widetilde{s}_{j}^{c}-1\right)\textrm{Gamma}\left(\widetilde{\rho}_{j}^{c};a,b\right)+\delta\left(\widetilde{s}_{j}^{c}+1\right)\textrm{Gamma}\left(\widetilde{\rho}_{j}^{c};\overline{a},\overline{b}\right)
\end{array}$\tabularnewline
\hline 
$\begin{array}{c}
u_{q}^{r}\\
u_{q}^{c}
\end{array}$ & $\begin{array}{c}
p\left(s_{q}^{r}\mid\overline{s}_{q}\right)\\
p\left(s_{q}^{c}\mid\overline{s}_{q}\right)
\end{array}$ & $\begin{array}{c}
p\left(s_{q}^{r}=1\mid\overline{s}_{q}=-1\right)=0,p\left(s_{q}^{r}=1\mid\overline{s}_{q}=1\right)=\lambda_{q}^{r}\\
p\left(s_{q}^{c}=1\mid\overline{s}_{q}=-1\right)=0,p\left(s_{q}^{c}=1\mid\overline{s}_{q}=1\right)=\lambda_{q}^{c}
\end{array}$\tabularnewline
\hline 
$\begin{array}{c}
\omega_{0}^{r}\\
\omega_{0}^{c}\\
\widetilde{\omega}_{j}^{r}
\end{array}$ & $\begin{array}{c}
p\left(s_{0}^{r}\right)\\
p\left(s_{0}^{c}\right)\\
p\left(\widetilde{s}_{j}^{c}\right)
\end{array}$ & $\begin{array}{c}
p\left(s_{0}^{r}=1\right)=\lambda_{0}^{r},p\left(s_{0}^{r}=-1\right)=1-\lambda_{0}^{r}\\
p\left(s_{0}^{c}=1\right)=\lambda_{0}^{c},p\left(s_{0}^{c}=-1\right)=1-\lambda_{0}^{c}\\
p\left(\widetilde{s}_{j}^{c}=1\right)=\widetilde{\lambda}_{j}^{c},p\left(\widetilde{s}_{j}^{c}=-1\right)=1-\widetilde{\lambda}_{j}^{c}
\end{array}$\tabularnewline
\hline 
\end{tabular}
\end{table*}

\subsection{Turbo-IF-VBI-E Step}

\begin{figure}[t]
\begin{centering}
\includegraphics[width=1\columnwidth]{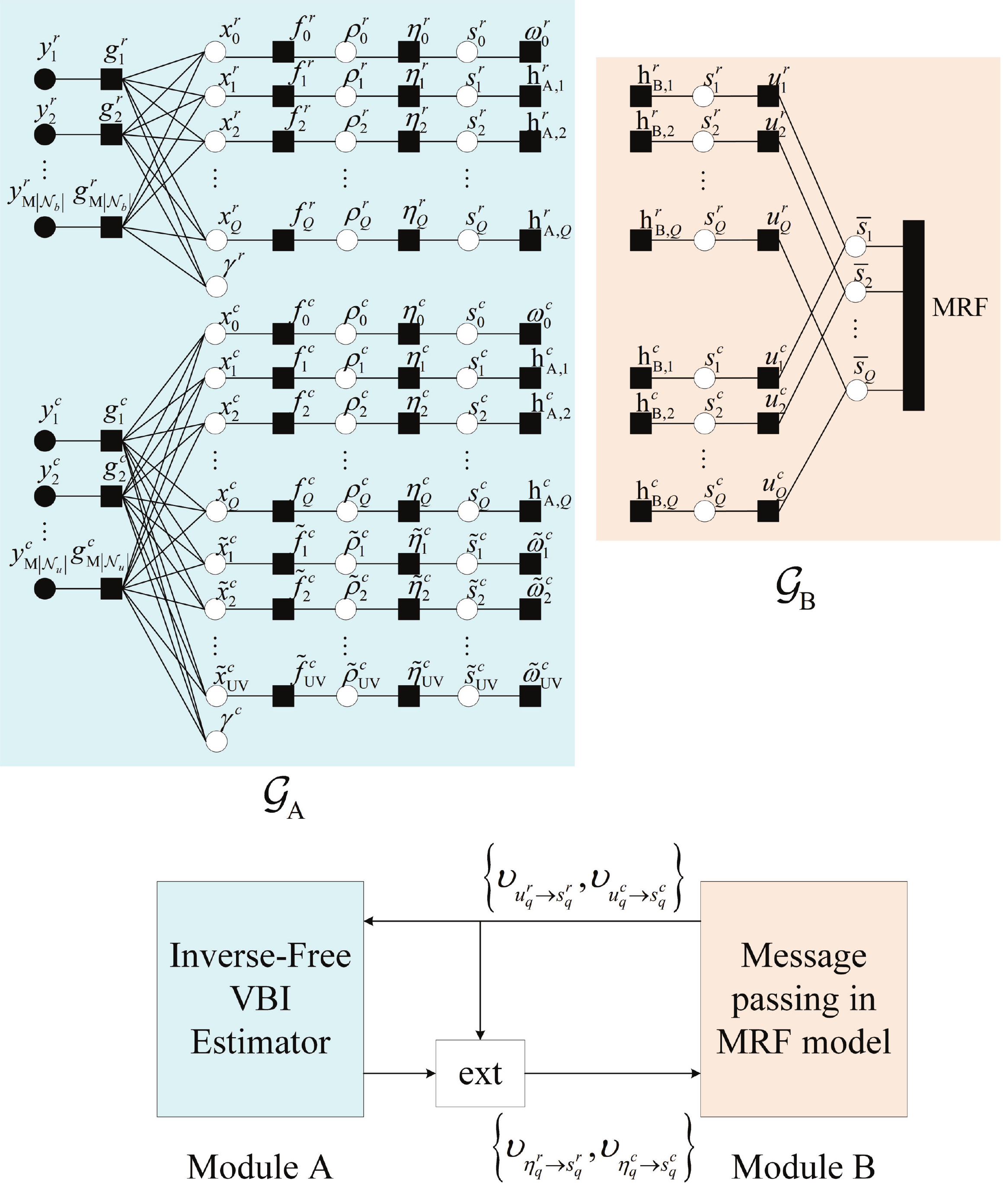}
\par\end{centering}
\caption{\label{fig:two_subgraph}Module A and Module B of the Turbo-IF-VBI-E
Step and messages flow between two modules.}
\end{figure}

The Turbo-IF-VBI-E Step is based on the turbo framework, which combines
the IF-VBI estimator with message passing, as shown in Fig. \ref{fig:Turbo-IF-VBI}.
The factor graph of the joint distribution $p\left(\boldsymbol{y},\boldsymbol{v},\overline{\boldsymbol{s}};\boldsymbol{\xi}\right)$
is illustrated in Fig. \ref{fig:factor graph}, where the expressions
of each factor node are listed in Table \ref{tab:Factors_Table}.
Since the factor graph has many loops, it is intractable to directly
perform Bayesian inference. For ease of implementation, we partition
the factor graph into two parts, denoted by $\mathcal{G_{A}}$ and
$\mathcal{G_{B}}$, respectively, where $\mathcal{G_{A}}$ models
the internal structure of the observation and $\mathcal{G_{B}}$ models
the internal structure of the support vectors. Correspondingly, we
introduce Module A and Module B to perform Bayesian inference over
$\mathcal{G_{A}}$ and $\mathcal{G_{B}}$, respectively. And the two
modules need to exchange messages with each other. Specifically, the
messages $\left\{ \upsilon_{\eta_{q}^{r}\rightarrow s_{q}^{r}},\upsilon_{\eta_{q}^{c}\rightarrow s_{q}^{c}}\right\} $
form the outputs of Module A and the inputs to Module B, while the
messages $\left\{ \upsilon_{u_{q}^{r}\rightarrow s_{q}^{r}},\upsilon_{u_{q}^{c}\rightarrow s_{q}^{c}}\right\} $form
the outputs of Module B and the inputs to Module A, as shown in Fig.
\ref{fig:two_subgraph}.

Before elaborating on Module A and Module B, we define two factor
nodes associated with turbo iteration,
\begin{equation}
\begin{aligned}\textrm{h}_{\textrm{A},q}^{r} & \triangleq\upsilon_{u_{q}^{r}\rightarrow s_{q}^{r}}\left(s_{q}^{r}\right) & \textrm{h}_{\textrm{A},q}^{c} & \triangleq\upsilon_{u_{q}^{c}\rightarrow s_{q}^{c}}\left(s_{q}^{c}\right),\\
\textrm{h}_{\textrm{B},q}^{r} & \triangleq\upsilon_{\eta_{q}^{r}\rightarrow s_{q}^{r}}\left(s_{q}^{r}\right) & \textrm{h}_{\textrm{B},q}^{c} & \triangleq\upsilon_{\eta_{q}^{c}\rightarrow s_{q}^{c}}\left(s_{q}^{c}\right),
\end{aligned}
\label{eq:turbo-iteration-factor}
\end{equation}
for $q=1,\ldots,Q$. For each turbo iteration, Module A treats $\textrm{h}_{\textrm{A},q}^{r}$
and $\textrm{h}_{\textrm{A},q}^{c}$ as the prior and performs the
IF-VBI estimator to calculate the approximate conditional posteriors.
Then, Module A passes extrinsic messages to Module B by subtracting
the prior information from posterior information, i.e.,
\begin{equation}
\begin{aligned}\upsilon_{\eta_{q}^{r}\rightarrow s_{q}^{r}}\left(s_{q}^{r}\right) & \propto q\left(s_{q}^{r}\right)/\textrm{h}_{\textrm{A},q}^{r},\forall q,\\
\upsilon_{\eta_{q}^{c}\rightarrow s_{q}^{c}}\left(s_{q}^{c}\right) & \propto q\left(s_{q}^{c}\right)/\textrm{h}_{\textrm{A},q}^{c},\forall q,
\end{aligned}
\label{eq:extricsic message}
\end{equation}
where $q\left(s_{q}^{r}\right)$ and $q\left(s_{q}^{c}\right)$ are
approximate marginal posteriors obtained in Module A. Similarly, Module
B performs message passing over $\mathcal{G_{B}}$ and passes the
extrinsic messages to Module A. The two modules iterate until they
reach a point of convergence.

\subsection{Inverse-Free VBI Estimator (Module A)}

We first give an overview of the variational Bayesian inference. For
convenience, let $\boldsymbol{v}^{l}$ denote an individual variable
in $\boldsymbol{v}$, such as $\boldsymbol{x},\boldsymbol{\rho},\boldsymbol{s},\gamma^{r},\gamma^{c}$.
Let $\mathcal{H}\triangleq\left\{ l\mid\forall\boldsymbol{v}^{l}\in\boldsymbol{v}\right\} $.
The posterior $p\left(\boldsymbol{v}\mid\boldsymbol{y};\boldsymbol{\xi}\right)$
is approximated by the product of some variational distributions,
\begin{equation}
p\left(\boldsymbol{v}\mid\boldsymbol{y};\boldsymbol{\xi}\right)\approx q\left(\boldsymbol{v}\right)=q\left(\boldsymbol{x}\right)q\left(\boldsymbol{\rho}\right)q\left(\boldsymbol{s}\right)q\left(\gamma^{r}\right)q\left(\gamma^{c}\right),
\end{equation}
where $q\left(\boldsymbol{v}^{l}\right),l\in\mathcal{H}$ are calculated
by maximizing the ELBO (equal to minimizing the KL-divergence). The
ELBO is given by
\begin{align}
L\left(q\right)= & \int_{\boldsymbol{v}}q\left(\boldsymbol{v}\right)\ln\dfrac{p\left(\boldsymbol{y},\boldsymbol{v};\boldsymbol{\xi}\right)}{q\left(\boldsymbol{v}\right)}\nonumber \\
= & \int_{\boldsymbol{v}}q\left(\boldsymbol{v}\right)\ln\Bigl(\dfrac{p\left(\boldsymbol{y}\mid\boldsymbol{x},\gamma^{r},\gamma^{c};\boldsymbol{\xi}\right)}{q\left(\boldsymbol{v}\right)}\nonumber \\
 & \hspace{10mm}p\left(\boldsymbol{x}\mid\boldsymbol{\rho}\right)p\left(\boldsymbol{\rho}\mid\boldsymbol{s}\right)p\left(\boldsymbol{s}\right)p\left(\gamma^{r}\right)p\left(\gamma^{c}\right)\Bigr).\label{eq:ELBO}
\end{align}
The authors in \cite{LiuAn_CE_Turbo_VBI} have proved that a stationary
solution, denoted by $q^{*}\left(\boldsymbol{v}\right)$, could be
found via alternately optimize each variational distribution $q\left(\boldsymbol{v}^{l}\right),\forall l\in\mathcal{H}$.
Specifically, for given $q\left(\boldsymbol{v}^{k}\right),\forall k\neq l$,
the optimal $q\left(\boldsymbol{v}^{l}\right)$ that maximizes the
ELBO can be obtained as
\begin{equation}
q\left(\boldsymbol{v}^{l}\right)=\frac{\exp\left(\left\langle \ln p\left(\boldsymbol{v},\boldsymbol{y}\right)\right\rangle _{\prod_{k\neq l}q\left(\boldsymbol{v}^{k}\right)}\right)}{\int_{\boldsymbol{v}^{l}}\exp\left(\left\langle \ln p\left(\boldsymbol{v},\boldsymbol{y}\right)\right\rangle _{\prod_{k\neq l}q\left(\boldsymbol{v}^{k}\right)}\right)},\label{eq:q(v)}
\end{equation}
where $\left\langle \cdot\right\rangle _{\prod_{k\neq l}q\left(\boldsymbol{v}^{k}\right)}$
denotes an expectation w.r.t the distributions $q\left(\boldsymbol{v}^{k}\right)$
for $k\neq l$. According to (\ref{eq:q(v)}), the update of $q\left(\boldsymbol{x}\right)$
is a Gaussian distribution with its mean and covariance matrix respectively
given by
\begin{equation}
\begin{aligned}\boldsymbol{\mu} & =\boldsymbol{\Sigma}\boldsymbol{\Phi}\left(\boldsymbol{\vartheta}\right)^{H}\left\langle \mathbf{\Gamma}\right\rangle \boldsymbol{y},\\
\boldsymbol{\Sigma} & =\left(\boldsymbol{\Phi}\left(\boldsymbol{\vartheta}\right)^{H}\left\langle \mathbf{\Gamma}\right\rangle \boldsymbol{\Phi}\left(\boldsymbol{\vartheta}\right)+\textrm{diag}\left(\left\langle \boldsymbol{\rho}\right\rangle \right)\right)^{-1},
\end{aligned}
\end{equation}
where $\boldsymbol{\Gamma}\triangleq\textrm{BlockDiag}\left(\gamma^{r}\mathbf{I}_{M\left|\mathcal{N}_{b}\right|},\gamma^{c}\mathbf{I}_{M\left|\mathcal{N}_{u}\right|}\right)$
is a diagonal matrix. Note that the update of $\boldsymbol{\Sigma}$
involves a matrix inverse, whose computational complexity is $\Theta\left(\left(1+Q+UV\right)^{3}\right)$.
Therefore, the algorithm is very time-consuming since $\left(1+Q+UV\right)$
is large.

To overcome this challenge, we follow the IF-SBL approach in \cite{Duan_IFSBL}
and avoid the matrix inverse via maximizing a relaxed ELBO instead.

Specifically, a lower bound of the likelihood function $p\left(\boldsymbol{y}\mid\boldsymbol{x},\gamma^{r},\gamma^{c};\boldsymbol{\xi}\right)$
can be obtained as 
\begin{align}
 & p\left(\boldsymbol{y}\mid\boldsymbol{x},\gamma^{r},\gamma^{c};\boldsymbol{\xi}\right)\nonumber \\
= & \dfrac{\det\left(\boldsymbol{\Gamma}\right)}{\pi^{M\left(\left|\mathcal{N}_{b}\right|+\left|\mathcal{N}_{u}\right|\right)}}\exp\left(-\left(\boldsymbol{y}-\boldsymbol{\Phi}\left(\boldsymbol{\vartheta}\right)\boldsymbol{x}\right)^{H}\boldsymbol{\Gamma}\left(\boldsymbol{y}-\boldsymbol{\Phi}\left(\boldsymbol{\vartheta}\right)\boldsymbol{x}\right)\right)\nonumber \\
\geq & \dfrac{\det\left(\boldsymbol{\Gamma}\right)}{\pi^{M\left(\left|\mathcal{N}_{b}\right|+\left|\mathcal{N}_{u}\right|\right)}}\exp\left(-g\left(\boldsymbol{x},\boldsymbol{w}\right)\right)\triangleq F\left(\boldsymbol{y},\boldsymbol{x},\boldsymbol{w},\gamma^{r},\gamma^{c};\boldsymbol{\xi}\right),\label{eq:lower_bound_of_likeli}
\end{align}
where the inequality in (\ref{eq:lower_bound_of_likeli}) follows
from Lemma 1 in \cite{Duan_IFSBL}, and
\begin{align}
g\left(\boldsymbol{x},\boldsymbol{w}\right)\triangleq & \left(\boldsymbol{y}-\boldsymbol{\Phi}\left(\boldsymbol{\vartheta}\right)\boldsymbol{w}\right)^{H}\boldsymbol{\Gamma}\left(\boldsymbol{y}-\boldsymbol{\Phi}\left(\boldsymbol{\vartheta}\right)\boldsymbol{w}\right)\nonumber \\
 & +2\mathfrak{Re}\left\{ \left(\boldsymbol{x}-\boldsymbol{w}\right)^{H}\boldsymbol{\Phi}\left(\boldsymbol{\vartheta}\right)^{H}\boldsymbol{\Gamma}\left(\boldsymbol{\Phi}\left(\boldsymbol{\vartheta}\right)\boldsymbol{w}-\boldsymbol{y}\right)\right\} \nonumber \\
 & +\left(\boldsymbol{x}-\boldsymbol{w}\right)^{H}\boldsymbol{\Gamma}\mathbf{T}\left(\boldsymbol{x}-\boldsymbol{w}\right).
\end{align}
Here $\mathbf{T}$ needs to satisfy $\mathbf{T}\succeq\boldsymbol{\Phi}\left(\boldsymbol{\vartheta}\right)^{H}\boldsymbol{\Phi}\left(\boldsymbol{\vartheta}\right)$.
And a good choice of $\mathbf{T}$ is
\begin{equation}
\mathbf{T}=\textrm{BlockDiag}\left(T^{r}\mathbf{I}_{M\left|\mathcal{N}_{b}\right|},T^{c}\mathbf{I}_{M\left|\mathcal{N}_{u}\right|}\right),\label{eq:T}
\end{equation}
with
\begin{align*}
T^{r} & \triangleq\textrm{\ensuremath{\lambda_{max}}}\left(\boldsymbol{\Phi}^{r}\left(\boldsymbol{\vartheta}\right)^{H}\boldsymbol{\Phi}^{r}\left(\boldsymbol{\vartheta}\right)\right),\\
T^{c} & \textrm{\ensuremath{\triangleq\lambda_{max}}}\left(\boldsymbol{\Phi}^{c}\left(\boldsymbol{\vartheta}\right)^{H}\boldsymbol{\Phi}^{c}\left(\boldsymbol{\vartheta}\right)\right),
\end{align*}
 where $\textrm{\ensuremath{\lambda_{max}}}\left(\cdot\right)$ denotes
the biggest eigenvalue of a matrix. In this case, $\mathbf{T}$ is
a diagonal matrix.

Substituting (\ref{eq:lower_bound_of_likeli}) into (\ref{eq:ELBO}),
we obtain a relaxed ELBO as
\begin{equation}
L\left(q\right)\geq\widetilde{L}\left(q,\boldsymbol{w}\right)\triangleq\int_{\boldsymbol{v}}q\left(\boldsymbol{v}\right)\ln\dfrac{G\left(\boldsymbol{y},\boldsymbol{v},\boldsymbol{w};\boldsymbol{\xi}\right)}{q\left(\boldsymbol{v}\right)},
\end{equation}
where
\begin{align}
 & G\left(\boldsymbol{y},\boldsymbol{v},\boldsymbol{w};\boldsymbol{\xi}\right)\\
= & F\left(\boldsymbol{y},\boldsymbol{x},\boldsymbol{w},\gamma^{r},\gamma^{c};\boldsymbol{\xi}\right)p\left(\boldsymbol{x}\mid\boldsymbol{\rho}\right)p\left(\boldsymbol{\rho}\mid\boldsymbol{s}\right)p\left(\boldsymbol{s}\right)p\left(\gamma^{r}\right)p\left(\gamma^{c}\right).\nonumber 
\end{align}
Now we maximize the relaxed ELBO $\widetilde{L}\left(q,\boldsymbol{w}\right)$
w.r.t $q\left(\boldsymbol{v}\right)$ and $\boldsymbol{w}$ based
on the EM method. Specifically, for given parameters $\boldsymbol{w}$,
optimize each variational distribution $q\left(\boldsymbol{v}^{l}\right),\forall l\in\mathcal{H}$
alternately. On the other hand, for given $q\left(\boldsymbol{v}\right)$,
maximize $\widetilde{L}\left(q,\boldsymbol{w}\right)$ w.r.t $\boldsymbol{w}$.

\subsubsection{Update of $q\left(\boldsymbol{v}\right)$}

Using (\ref{eq:q(v)}) and ignoring the terms that are not depend
on $\boldsymbol{x}$, $q\left(\boldsymbol{x}\right)$ can be derived
as
\begin{align}
\ln q\left(\boldsymbol{x}\right)\propto & \ln F\left(\boldsymbol{y},\boldsymbol{x},\boldsymbol{w},\gamma^{r},\gamma^{c};\boldsymbol{\xi}\right)+\ln p\left(\boldsymbol{x}\mid\boldsymbol{\rho}\right)\nonumber \\
\propto & -\boldsymbol{x}^{H}\left(\left\langle \boldsymbol{\Gamma}\right\rangle \mathbf{T}+\textrm{diag}\left(\left\langle \boldsymbol{\rho}\right\rangle \right)\right)\boldsymbol{x}\nonumber \\
 & +2\mathfrak{Re}\left\{ \boldsymbol{x}^{H}\boldsymbol{\Phi}\left(\boldsymbol{\vartheta}\right)^{H}\left\langle \boldsymbol{\Gamma}\right\rangle \left(\boldsymbol{y}-\boldsymbol{\Phi}\left(\boldsymbol{\vartheta}\right)\boldsymbol{w}\right)\right\} \nonumber \\
 & +2\mathfrak{Re}\left\{ \boldsymbol{x}^{H}\left\langle \boldsymbol{\Gamma}\right\rangle \mathbf{T}\boldsymbol{w}\right\} \nonumber \\
\propto & \ln\mathcal{CN}\left(\boldsymbol{x};\boldsymbol{\mu},\boldsymbol{\Sigma}\right).\label{eq:q(x)}
\end{align}
where the posterior mean and covariance matrix are respectively given
by
\begin{equation}
\begin{aligned}\boldsymbol{\mu} & =\boldsymbol{\Sigma}\left(\boldsymbol{\Phi}\left(\boldsymbol{\vartheta}\right)^{H}\left\langle \boldsymbol{\Gamma}\right\rangle \left(\boldsymbol{y}-\boldsymbol{\Phi}\left(\boldsymbol{\vartheta}\right)\boldsymbol{w}\right)+\left\langle \boldsymbol{\Gamma}\right\rangle \mathbf{T}\boldsymbol{w}\right),\\
\boldsymbol{\Sigma} & =\left(\left\langle \boldsymbol{\Gamma}\right\rangle \mathbf{T}+\textrm{diag}\left(\left\langle \boldsymbol{\rho}\right\rangle \right)\right)^{-1}.
\end{aligned}
\label{eq:x_post}
\end{equation}
Note that $\boldsymbol{\Sigma}$ is calculated by a diagonal matrix
inverse and $\boldsymbol{\mu}$ is computed by the matrix-vector product.
Therefore, the computational complexity of $q\left(\boldsymbol{x}\right)$
is significantly reduced.

The update of $q\left(\boldsymbol{\rho}\right)$, $q\left(\boldsymbol{s}\right)$,
$q\left(\gamma^{r}\right)$, and $q\left(\gamma^{c}\right)$ can be
derived in the same way. Please refer to the Turbo-VBI algorithm in
\cite{LiuAn_CE_Turbo_VBI} for the expressions of these variational
distributions.

\subsubsection{Update of $\boldsymbol{w}$}

Submitting $q\left(\boldsymbol{v};\boldsymbol{w}^{\textrm{old}}\right)$
into $\widetilde{L}\left(q,\boldsymbol{w}\right)$, an estimate of
$\boldsymbol{w}$ can be obtained as
\begin{equation}
\boldsymbol{w}^{\textrm{new}}=\arg\max_{\boldsymbol{w}}\left\langle \ln G\left(\boldsymbol{y},\boldsymbol{v},\boldsymbol{w};\boldsymbol{\xi}\right)\right\rangle _{q\left(\boldsymbol{v};\boldsymbol{w}^{\textrm{old}}\right)}.
\end{equation}
The gradient of $\left\langle \ln G\left(\boldsymbol{y},\boldsymbol{v},\boldsymbol{w};\boldsymbol{\xi}\right)\right\rangle _{q\left(\boldsymbol{v};\boldsymbol{w}^{\textrm{old}}\right)}$
w.r.t $\boldsymbol{w}$ is given by
\begin{align}
 & \frac{\partial\left\langle \ln G\left(\boldsymbol{y},\boldsymbol{v},\boldsymbol{w};\boldsymbol{\xi}\right)\right\rangle _{q\left(\boldsymbol{v};\boldsymbol{w}^{\textrm{old}}\right)}}{\partial\boldsymbol{w}}\nonumber \\
= & \left\langle \boldsymbol{\Gamma}\right\rangle \left(\mathbf{T}-\boldsymbol{\Phi}\left(\boldsymbol{\vartheta}\right)^{H}\boldsymbol{\Phi}\left(\boldsymbol{\vartheta}\right)\right)\left(\boldsymbol{\mu}-\boldsymbol{w}\right).
\end{align}
By setting the gradient to zero, we have
\begin{equation}
\boldsymbol{w}^{\textrm{new}}=\boldsymbol{\mu}.\label{eq:w_new}
\end{equation}

\subsection{Message Passing in MRF (Module B)}

In $\mathcal{G}_{B}$, the sub factor graphs associated with $\boldsymbol{s}^{r}$
and $\boldsymbol{s}^{c}$ are coupled together via the variable nodes
$\left\{ \overline{s}_{q}\right\} $. Therefore, they exchange messages
to obtain more accurate estimates for $\boldsymbol{s}^{r}$ and $\boldsymbol{s}^{c}$.
In other words, the sensing and communication functionalities assist
each other when performing message passing. We follow the sum-product
rule to derive the message passing algorithm over $\mathcal{G}_{B}$.
To simplify the notation, $\pi_{q}^{r}$, $\pi_{q}^{c}$, $\pi_{q}^{r,in}$,
and $\pi_{q}^{c,in}$ are used to abbreviate $\textrm{h}_{\textrm{A},q}^{r}\left(s_{q}^{r}=1\right)$,
$\textrm{h}_{\textrm{A},q}^{c}\left(s_{q}^{c}=1\right)$, $\textrm{h}_{\textrm{B},q}^{r}\left(s_{q}^{r}=1\right)$,
and $\textrm{h}_{\textrm{B},q}^{c}\left(s_{q}^{c}=1\right)$, respectively,
for $q=1,\ldots,Q$. The message from the factor nodes $u_{q}^{r}$
and $u_{q}^{c}$ to the variable node $\overline{s}_{q}$ are respectively
given by
\begin{equation}
\begin{aligned}\nu_{u_{q}^{r}\rightarrow\overline{s}_{q}} & \propto\sum_{s_{q}^{r}}\nu_{s_{q}^{r}\rightarrow u_{q}^{r}}\times u_{q}^{r}\left(s_{q}^{r},\overline{s}_{q}\right)\\
 & \propto\pi_{u_{q}^{r}\rightarrow\overline{s}_{q}}\delta\left(\overline{s}_{q}-1\right)+\left(1-\pi_{u_{q}^{r}\rightarrow\overline{s}_{q}}\right)\delta\left(\overline{s}_{q}+1\right),\\
\nu_{u_{q}^{c}\rightarrow\overline{s}_{q}} & \propto\sum_{s_{q}^{c}}\nu_{s_{q}^{c}\rightarrow u_{q}^{c}}\times u_{q}^{c}\left(s_{q}^{c},\overline{s}_{q}\right)\\
 & \propto\pi_{u_{q}^{c}\rightarrow\overline{s}_{q}}\delta\left(\overline{s}_{q}-1\right)+\left(1-\pi_{u_{q}^{c}\rightarrow\overline{s}_{q}}\right)\delta\left(\overline{s}_{q}+1\right),
\end{aligned}
\label{eq:message1}
\end{equation}
where 
\begin{align*}
\pi_{u_{q}^{r}\rightarrow\overline{s}_{q}} & =\left(1+\tfrac{1-\pi_{q}^{r,in}}{1+2\lambda_{q}^{r}\pi_{q}^{r,in}-\lambda_{q}^{r}-\pi_{q}^{r,in}}\right)^{-1},\\
\pi_{u_{q}^{c}\rightarrow\overline{s}_{q}} & =\left(1+\tfrac{1-\pi_{q}^{c,in}}{1+2\lambda_{q}^{c}\pi_{q}^{c,in}-\lambda_{q}^{c}-\pi_{q}^{c,in}}\right)^{-1}.
\end{align*}

Consider the variable node $\overline{s}_{q}$ and define the index
set of its neighbors as $\mathcal{N}_{q}\triangleq\left\{ q_{l},q_{r},q_{t},q_{b}\right\} $,
where the left, right, top, and bottom neighbors are $\overline{s}_{q_{l}}\triangleq\overline{s}_{q-H}$,
$\overline{s}_{q_{r}}\triangleq\overline{s}_{q+H}$, $\overline{s}_{q_{t}}\triangleq\overline{s}_{q-1}$,
and $\overline{s}_{q_{b}}\triangleq\overline{s}_{q+1}$, respectively.
Then the input messages of $\overline{s}_{q}$ from the left, right,
top, and bottom, denoted by $\nu_{q}^{l}$, $\nu_{q}^{r}$, $\nu_{q}^{t}$,
and $\nu_{q}^{b}$, are Bernoulli distributions, where $\nu_{q}^{l}$
can be calculated as
\begin{align}
\nu_{q}^{l} & \propto\sum_{\overline{s}_{q_{l}}}\nu_{u_{q_{l}}^{r}\rightarrow\overline{s}_{q_{l}}}\nu_{u_{q_{l}}^{c}\rightarrow\overline{s}_{q_{l}}}\prod_{k\in\left\{ l,t,b\right\} }\nu_{q_{l}}^{k}\psi\left(\overline{s}_{q_{l}}\right)\varphi\left(\overline{s}_{q},\overline{s}_{q_{l}}\right)\nonumber \\
 & \propto\kappa_{q}^{l}\delta\left(\overline{s}_{q}-1\right)+(1-\kappa_{q}^{l})\delta\left(\overline{s}_{q}+1\right),\label{eq:message2}
\end{align}
where $\kappa_{q}^{l}$ is given in (\ref{eq:kql}) at the top of
the next page.
\begin{figure*}[!t]
\begin{equation}
\kappa_{q}^{l}=\dfrac{\pi_{u_{q_{l}}^{r}\rightarrow\overline{s}_{q_{l}}}\pi_{u_{q_{l}}^{c}\rightarrow\overline{s}_{q_{l}}}\prod_{k\in\left\{ l,t,b\right\} }\kappa_{q_{l}}^{k}e^{-\alpha_{q_{l}}+\beta_{q,q_{l}}}+\left(1-\pi_{u_{q_{l}}^{r}\rightarrow\overline{s}_{q_{l}}}\right)\left(1-\pi_{u_{q_{l}}^{c}\rightarrow\overline{s}_{q_{l}}}\right)\prod_{k\in\left\{ l,t,b\right\} }\left(1-\kappa_{q_{l}}^{k}\right)e^{\alpha_{q_{l}}-\beta_{q,q_{l}}}}{\left(e^{\beta_{q,q_{l}}}+e^{-\beta_{q,q_{l}}}\right)\left(\pi_{u_{q_{l}}^{r}\rightarrow\overline{s}_{q_{l}}}\pi_{u_{q_{l}}^{c}\rightarrow\overline{s}_{q_{l}}}e^{-\alpha_{q_{l}}}\prod_{k\in\left\{ l,t,b\right\} }\kappa_{q_{l}}^{k}+\left(1-\pi_{u_{q_{l}}^{r}\rightarrow\overline{s}_{q_{l}}}\right)\left(1-\pi_{u_{q_{l}}^{c}\rightarrow\overline{s}_{q_{l}}}\right)e^{\alpha_{q_{l}}}\prod_{k\in\left\{ l,t,b\right\} }\left(1-\kappa_{q_{l}}^{k}\right)\right)}.\label{eq:kql}
\end{equation}
\end{figure*}

The other input messages of $\overline{s}_{q}$, i.e., $\nu_{q}^{r}$,
$\nu_{q}^{t}$ and $\nu_{q}^{b}$, have a similar form to $\nu_{q}^{l}$.

The message from the variable node $\overline{s}_{q}$ to the factor
node $u_{q}^{r}$ is given by
\begin{align}
\nu_{\overline{s}_{q}\rightarrow u_{q}^{r}} & \propto\nu_{u_{q}^{c}\rightarrow\overline{s}_{q}}\Pi_{k\in\left\{ l,r,t,b\right\} }\nu_{q}^{k}\psi\left(\overline{s}_{q}\right)\label{eq:message3}\\
 & \propto\pi_{\overline{s}_{q}\rightarrow u_{q}^{r}}\delta\left(\overline{s}_{q}-1\right)+\left(1-\pi_{\overline{s}_{q}\rightarrow u_{q}^{r}}\right)\delta\left(\overline{s}_{q}+1\right),\nonumber 
\end{align}
where $\pi_{\overline{s}_{q}\rightarrow u_{q}^{r}}$ is given in (\ref{eq:pi_out})
at the top of the next page.
\begin{figure*}[!t]
\begin{equation}
\pi_{\overline{s}_{q}\rightarrow u_{q}^{r}}=\frac{e^{-\alpha_{q}}\pi_{u_{q}^{c}\rightarrow\overline{s}_{q}}\Pi_{k\in\left\{ l,r,t,b\right\} }\kappa_{q}^{k}}{e^{-\alpha_{q}}\pi_{u_{q}^{c}\rightarrow\overline{s}_{q}}\Pi_{k\in\left\{ l,r,t,b\right\} }\kappa_{q}^{k}+e^{\alpha_{q}}\left(1-\pi_{u_{q}^{c}\rightarrow\overline{s}_{q}}\right)\Pi_{k\in\left\{ l,r,t,b\right\} }\left(1-\kappa_{q}^{k}\right)}.\label{eq:pi_out}
\end{equation}

\rule[0.5ex]{1\textwidth}{1pt}
\end{figure*}

The message from the factor node $u_{q}^{r}$ back to the variable
node $s_{q}^{r}$ is given by
\begin{align}
\nu_{u_{q}^{r}\rightarrow s_{q}^{r}} & \propto\sum_{\overline{s}_{q}}\nu_{\overline{s}_{q}\rightarrow u_{q}^{r}}\times u_{q}^{r}\left(s_{q}^{r},\overline{s}_{q}\right)\nonumber \\
 & \propto\pi_{q}^{r}\delta\left(s_{q}^{r}-1\right)+\left(1-\pi_{q}^{r}\right)\delta\left(s_{q}^{r}+1\right),\label{eq:message4}
\end{align}
where $\pi_{q}^{r}=\pi_{\overline{s}_{q}\rightarrow u_{q}^{r}}\lambda_{q}^{r}$.
Similarly, the message from the factor node $u_{q}^{c}$ back to the
variable node $s_{q}^{c}$ is given by
\begin{equation}
\nu_{u_{q}^{c}\rightarrow s_{q}^{c}}=\pi_{q}^{c}\delta\left(s_{q}^{c}-1\right)+\left(1-\pi_{q}^{c}\right)\delta\left(s_{q}^{c}+1\right),\label{eq:message5}
\end{equation}
where $\pi_{q}^{c}=\pi_{\overline{s}_{q}\rightarrow u_{q}^{c}}\lambda_{q}^{c}$.

The approximate marginal posterior $q\left(\overline{\boldsymbol{s}}\mid\boldsymbol{y};\boldsymbol{\xi}\right)$
can be obtained as
\begin{align}
q\left(\overline{\boldsymbol{s}}\mid\boldsymbol{y};\boldsymbol{\xi}\right) & \propto\prod_{q}\nu_{u_{q}^{r}\rightarrow\overline{s}_{q}}\times\nu_{\overline{s}_{q}\rightarrow u_{q}^{r}}\label{eq:q(s_bar)}\\
 & \propto\prod_{q}\left(\pi_{\overline{s}_{q}}\delta\left(\overline{s}_{q}-1\right)+\left(1-\pi_{\overline{s}_{q}}\right)\delta\left(\overline{s}_{q}+1\right)\right),\nonumber 
\end{align}
where $\pi_{\overline{s}_{q}}=\frac{\pi_{u_{q}^{r}\rightarrow\overline{s}_{q}}\pi_{\overline{s}_{q}\rightarrow u_{q}^{r}}}{\pi_{u_{q}^{r}\rightarrow\overline{s}_{q}}\pi_{\overline{s}_{q}\rightarrow u_{q}^{r}}+\left(1-\pi_{u_{q}^{r}\rightarrow\overline{s}_{q}}\right)\left(\pi_{\overline{s}_{q}\rightarrow u_{q}^{r}}\right)},\forall q$.

\subsection{Turbo-IF-VBI-M Step\label{subsec:Turbo-IF-VBI-M-Step}}

Since there is no close-form expression of $\ln p\left(\boldsymbol{y},\boldsymbol{\xi}\right)$,
it is challenging to directly solve the maximization problem in (\ref{eq:M-step}).
To get around this problem, one common solution is to construct a
surrogate function of $\ln p\left(\boldsymbol{y},\boldsymbol{\xi}\right)$
and maximize the surrogate function with respect to $\boldsymbol{\xi}$.
Specifically, in the $t\textrm{-th}$ iteration, the surrogate function
inspired by the EM method is given by
\begin{align}
Q\left(\boldsymbol{\xi};\boldsymbol{\xi}^{\left(t\right)}\right) & =\sum_{\overline{\boldsymbol{s}}}\int_{\boldsymbol{v}}q^{\left(t\right)}\left(\boldsymbol{v},\overline{\boldsymbol{s}}\right)\ln\frac{p\left(\boldsymbol{y},\boldsymbol{v},\overline{\boldsymbol{s}};\boldsymbol{\xi}\right)}{q^{\left(t\right)}\left(\boldsymbol{v},\overline{\boldsymbol{s}}\right)}+\ln p\left(\boldsymbol{\vartheta}\right)\nonumber \\
 & =Q_{\boldsymbol{\vartheta}}\left(\boldsymbol{\vartheta};\boldsymbol{\xi}^{\left(t\right)}\right)+Q_{\boldsymbol{\zeta}}\left(\boldsymbol{\zeta};\boldsymbol{\xi}^{\left(t\right)}\right)+C,\label{eq:func_Q}
\end{align}
where
\begin{align}
Q_{\boldsymbol{\vartheta}}\left(\boldsymbol{\vartheta};\boldsymbol{\xi}^{\left(t\right)}\right)= & -\left(\boldsymbol{y}-\boldsymbol{\Phi}\left(\boldsymbol{\vartheta}\right)\boldsymbol{\mu}^{\left(t\right)}\right)^{H}\boldsymbol{\Gamma}\left(\boldsymbol{y}-\boldsymbol{\Phi}\left(\boldsymbol{\vartheta}\right)\boldsymbol{\mu}^{\left(t\right)}\right)\nonumber \\
 & -\textrm{tr}\left(\boldsymbol{\Gamma}\boldsymbol{\Phi}\left(\boldsymbol{\vartheta}\right)\boldsymbol{\Sigma}^{\left(t\right)}\boldsymbol{\Phi}\left(\boldsymbol{\vartheta}\right)^{H}\right)+\ln p\left(\boldsymbol{\vartheta}\right),\nonumber \\
Q_{\boldsymbol{\zeta}}\left(\boldsymbol{\zeta};\boldsymbol{\xi}^{\left(t\right)}\right)= & \mathbb{E}_{q\left(\overline{\boldsymbol{s}}\mid\boldsymbol{y};\boldsymbol{\xi}^{\left(t\right)}\right)}\left\{ \ln p\left(\overline{\boldsymbol{s}};\boldsymbol{\zeta}\right)\right\} ,\label{eq:func_Q2}
\end{align}
where $\boldsymbol{\mu}^{\left(t\right)}$ and $\boldsymbol{\Sigma}^{\left(t\right)}$
are approximate posterior mean and covariance matrix of $\boldsymbol{x}$
obtained in (\ref{eq:x_post}), $q^{\left(t\right)}\left(\boldsymbol{v},\overline{\boldsymbol{s}}\right)\triangleq q\left(\boldsymbol{v}\mid\boldsymbol{y};\boldsymbol{\xi}^{\left(t\right)}\right)q\left(\overline{\boldsymbol{s}}\mid\boldsymbol{y};\boldsymbol{\xi}^{\left(t\right)}\right)$,
and $C$ is a constant. At the current iterate $\boldsymbol{\xi}^{\left(t\right)}$,
the surrogate function and its gradient satisfy the following properties:
\begin{subequations}
\begin{align}
Q\left(\boldsymbol{\xi};\boldsymbol{\xi}^{\left(t\right)}\right) & \leq\ln p\left(\boldsymbol{y},\boldsymbol{\xi}\right),\forall\boldsymbol{\xi}\\
Q\left(\boldsymbol{\xi}^{\left(t\right)};\boldsymbol{\xi}^{\left(t\right)}\right) & =\ln p\left(\boldsymbol{y},\boldsymbol{\xi}^{\left(t\right)}\right),\\
\frac{\partial Q\left(\boldsymbol{\xi}^{\left(t\right)};\boldsymbol{\xi}^{\left(t\right)}\right)}{\partial\boldsymbol{\xi}} & =\frac{\partial\ln p\left(\boldsymbol{y},\boldsymbol{\xi}^{\left(t\right)}\right)}{\partial\boldsymbol{\xi}}.
\end{align}
\end{subequations}
Maximizing $Q\left(\boldsymbol{\xi};\boldsymbol{\xi}^{\left(t\right)}\right)$
is equal to solve the following two subproblems:
\begin{subequations}
\begin{align}
\boldsymbol{\vartheta}^{\left(t+1\right)} & =\underset{\boldsymbol{\vartheta}}{\arg\max}Q_{\boldsymbol{\vartheta}}\left(\boldsymbol{\vartheta};\boldsymbol{\xi}^{\left(t\right)}\right),\label{eq:subproblem_Q1}\\
\boldsymbol{\zeta}^{\left(t+1\right)} & =\underset{\boldsymbol{\zeta}}{\arg\max}Q_{\boldsymbol{\zeta}}\left(\boldsymbol{\zeta};\boldsymbol{\xi}^{\left(t\right)}\right).\label{eq:subproblem_Q2}
\end{align}
\end{subequations}

\subsubsection{Update of $\boldsymbol{\vartheta}$}

It is difficult to find the global optimal solution to the subproblem
in (\ref{eq:subproblem_Q1}) because the function $Q_{\boldsymbol{\vartheta}}\left(\boldsymbol{\vartheta};\boldsymbol{\xi}^{\left(t\right)}\right)$
is non-convex. Using the gradient ascent method, we have
\begin{align}
\boldsymbol{r}^{\left(t+1\right)} & =\boldsymbol{r}^{\left(t\right)}+\varepsilon_{r}^{\left(t\right)}\frac{\partial Q_{\boldsymbol{\vartheta}}\left(\boldsymbol{r}^{\left(t\right)},\boldsymbol{p}_{u}^{\left(t\right)},\tau_{o}^{\left(t\right)};\boldsymbol{\xi}^{\left(t\right)}\right)}{\partial\boldsymbol{r}},\label{eq:update_r}\\
\boldsymbol{p}_{u}^{\left(t+1\right)} & =\boldsymbol{p}_{u}^{\left(t\right)}+\varepsilon_{p}^{\left(t\right)}\frac{\partial Q_{\boldsymbol{\vartheta}}\left(\boldsymbol{r}^{\left(t+1\right)},\boldsymbol{p}_{u}^{\left(t\right)},\tau_{o}^{\left(t\right)};\boldsymbol{\xi}^{\left(t\right)}\right)}{\partial\boldsymbol{p}_{u}},\label{eq:update_p}\\
\tau_{o}^{\left(t+1\right)} & =\tau_{o}^{\left(t\right)}+\varepsilon_{\tau}^{\left(t\right)}\frac{\partial Q_{\boldsymbol{\vartheta}}\left(\boldsymbol{r}^{\left(t+1\right)},\boldsymbol{p}_{u}^{\left(t+1\right)},\tau_{o}^{\left(t\right)};\boldsymbol{\xi}^{\left(t\right)}\right)}{\partial\tau_{o}},\label{eq:update_t}
\end{align}
where $\varepsilon_{r}^{\left(t\right)}$, $\varepsilon_{p}^{\left(t\right)}$,
and $\varepsilon_{\tau}^{\left(t\right)}$ are step sizes determined
by the Armijo rule.

\subsubsection{Update of $\boldsymbol{\zeta}$}

Recalling (\ref{eq:func_Q2}) and (\ref{eq:p(s)}), we find that the
partition function $Z\left(\boldsymbol{\zeta}\right)$ is computationally
intractable due to its exponential complexity. Therefore, it is very
difficult to directly apply the gradient ascent method to solve the
subproblem in (\ref{eq:subproblem_Q2}). To make the subproblem solvable,
a pseudo-likelihood function is introduced to approximate the MRF
prior,
\begin{align}
\textrm{PL}\left(\overline{\boldsymbol{s}};\boldsymbol{\zeta}\right) & =\prod_{q\in\mathcal{S}\setminus\partial\mathcal{S}}p\left(\overline{s}_{q}\mid\overline{s}_{q-H},\overline{s}_{q+H},\overline{s}_{q-1},\overline{s}_{q+1}\right)\\
 & =\prod_{q\in\mathcal{S}\setminus\partial\mathcal{S}}\frac{\exp\left(-\alpha_{q}\overline{s}_{q}+\sum_{i\in\mathcal{N}_{q}}\beta_{iq}\overline{s}_{i}\overline{s}_{q}\right)}{\sum_{\overline{s}_{q}}\exp\left(-\alpha_{q}\overline{s}_{q}+\sum_{i\in\mathcal{N}_{q}}\beta_{iq}\overline{s}_{i}\overline{s}_{q}\right)},\nonumber 
\end{align}
where $\mathcal{S}\triangleq\left\{ 1,\ldots,Q\right\} $ is the index
set of variable nodes in the MRF model and $\partial\mathcal{S}$
is the index set of nodes at the boundaries of $\mathcal{S}$. The
authors in \cite{Geman_MRF} proved the consistency of the maximum
pseudo-likelihood estimate for large $H$ and $W$. In other words,
the global optimal solution of maximizing $\ln\textrm{PL}\left(\overline{\boldsymbol{s}};\boldsymbol{\zeta}\right)$
converges to the global optimal of maximizing $\ln p\left(\boldsymbol{s};\boldsymbol{\zeta}\right)$
as $H,W\rightarrow\infty$. Therefore, for large $H$ and $W$, by
replacing the likelihood $p\left(\overline{\boldsymbol{s}};\boldsymbol{\zeta}\right)$
in $Q_{\boldsymbol{\zeta}}\left(\boldsymbol{\zeta};\boldsymbol{\xi}^{\left(t\right)}\right)$
with the pseudo-likelihood, we obtain a good approximation to the
subproblem as
\begin{align}
\boldsymbol{\zeta}^{\left(t+1\right)} & =\underset{\boldsymbol{\zeta}}{\arg\max}\widetilde{Q}_{\boldsymbol{\zeta}}\left(\boldsymbol{\zeta};\boldsymbol{\xi}^{\left(t\right)}\right)\nonumber \\
 & =\underset{\boldsymbol{\zeta}}{\arg\max}\mathbb{E}_{q\left(\overline{\boldsymbol{s}}\mid\boldsymbol{y};\boldsymbol{\xi}^{\left(t\right)}\right)}\left\{ \ln\textrm{PL}\left(\overline{\boldsymbol{s}};\boldsymbol{\zeta}\right)\right\} .
\end{align}
The gradients of $\widetilde{Q}_{\boldsymbol{\zeta}}\left(\boldsymbol{\zeta};\boldsymbol{\xi}^{\left(t\right)}\right)$
w.r.t. $\alpha_{q}$ and $\beta_{pq}$ are respectively given in (\ref{eq:gradient_alpha})
and (\ref{eq:gradient_beta}) at the top of the next page, where $p,q\in\mathcal{S}\setminus\partial\mathcal{S}$
and $p\in\mathcal{N}_{q}$.
\begin{figure*}[!t]
\begin{equation}
\frac{\partial\widetilde{Q}_{\boldsymbol{\zeta}}\left(\boldsymbol{\zeta};\boldsymbol{\xi}^{\left(t\right)}\right)}{\partial\alpha_{q}}=\mathbb{E}_{q\left(\overline{\boldsymbol{s}}\mid\boldsymbol{y};\boldsymbol{\xi}^{\left(t\right)}\right)}\left\{ -\overline{s}_{q}+\frac{\exp\left(2\sum_{i\in\mathcal{N}_{q}}\beta_{iq}\overline{s}_{i}\right)-\exp\left(2\alpha_{q}\right)}{\exp\left(2\sum_{i\in\mathcal{N}_{q}}\beta_{iq}\overline{s}_{i}\right)+\exp\left(2\alpha_{q}\right)}\right\} ,\label{eq:gradient_alpha}
\end{equation}
\begin{equation}
\frac{\partial\widetilde{Q}_{\boldsymbol{\zeta}}\left(\boldsymbol{\zeta};\boldsymbol{\xi}^{\left(t\right)}\right)}{\partial\beta_{pq}}=\mathbb{E}_{q\left(\overline{\boldsymbol{s}}\mid\boldsymbol{y};\boldsymbol{\xi}^{\left(t\right)}\right)}\left\{ 2\overline{s}_{p}\overline{s}_{q}-\overline{s}_{p}\frac{\exp\left(2\sum_{i\in\mathcal{N}_{q}}\beta_{iq}\overline{s}_{i}\right)-\exp\left(2\alpha_{q}\right)}{\exp\left(2\sum_{i\in\mathcal{N}_{q}}\beta_{iq}\overline{s}_{i}\right)+\exp\left(2\alpha_{q}\right)}-\overline{s}_{q}\frac{\exp\left(2\sum_{i\in\mathcal{N}_{p}}\beta_{ip}\overline{s}_{i}\right)-\exp\left(2\alpha_{p}\right)}{\exp\left(2\sum_{i\in\mathcal{N}_{p}}\beta_{ip}\overline{s}_{i}\right)+\exp\left(2\alpha_{p}\right)}\right\} ,\label{eq:gradient_beta}
\end{equation}

\rule[0.5ex]{1\textwidth}{1pt}
\end{figure*}
 This time, the gradients can be directly calculated, so the gradient
ascent method is feasible without resorting to the time-consuming
Monte Carlo method.

We follow the gradient ascent approach, which leads to the following
update:
\begin{align}
\alpha_{q}^{\left(t+1\right)} & =\alpha_{q}^{\left(t\right)}+\varepsilon_{\alpha}\frac{\partial\widetilde{Q}_{\boldsymbol{\zeta}}\left(\boldsymbol{\zeta};\boldsymbol{\xi}^{\left(t\right)}\right)}{\partial\alpha_{q}}\mid_{\boldsymbol{\zeta}=\boldsymbol{\zeta}^{\left(t\right)}},\label{eq:update_alpha_2}\\
\beta_{pq}^{\left(t+1\right)} & =\beta_{pq}^{\left(t\right)}+\varepsilon_{\beta}\frac{\partial\widetilde{Q}_{\boldsymbol{\zeta}}\left(\boldsymbol{\zeta};\boldsymbol{\xi}^{\left(t\right)}\right)}{\partial\beta_{pq}}\mid_{\boldsymbol{\zeta}=\boldsymbol{\zeta}^{\left(t\right)}},\label{eq:update_beta_2}
\end{align}
where $\varepsilon_{\alpha}$ and $\varepsilon_{\beta}$ are step
sizes determined by the Armijo rule.

The complete Turbo-IF-VBI algorithm is summarized in Algorithm \ref{Turbo-IF-VBI}.

\begin{algorithm}[t]
\begin{singlespace}
{\small{}\caption{\label{Turbo-IF-VBI}Turbo-IF-VBI algorithm}
}{\small\par}

\textbf{Input:} $\boldsymbol{y}$, $\boldsymbol{\Phi}\left(\boldsymbol{\vartheta}\right)$,
$p\left(\boldsymbol{\vartheta}\right)$, iteration numbers $I_{in}$
and $I_{out}$, threshold $\epsilon$.

\textbf{Output:} $\boldsymbol{\xi}^{*}$, $\boldsymbol{x}^{*}$, and
$\boldsymbol{s}^{*}$.

\begin{algorithmic}[1]

\FOR{${\color{blue}{\color{black}t=1,\cdots,I_{out}}}$}

\STATE \textbf{Turbo-IF-VBI-E Step:}

\STATE \textbf{\%Module A: IF-VBI Estimator}

\STATE Initialize $i_{in}=1$, $\boldsymbol{\xi}$, $\boldsymbol{\pi}$,
$\mathbf{T}$, and $\boldsymbol{w}=\boldsymbol{\mu}^{\left(t-1\right)}$,
where $\boldsymbol{\mu}^{\left(0\right)}\triangleq\boldsymbol{\Phi}\left(\boldsymbol{\vartheta}\right)^{H}\boldsymbol{y}$.

\WHILE{not converge and $i_{in}\leq I_{in}$}

\STATE Update $q\left(\boldsymbol{v}\mid\boldsymbol{y};\boldsymbol{\xi}^{\left(t\right)}\right)$,
using (\ref{eq:x_post}).

\STATE Update the parameter $\boldsymbol{w}$, using (\ref{eq:w_new}).

\STATE $i_{in}=i_{in}+1$.

\ENDWHILE

\STATE Calculate the extrinsic information based on (\ref{eq:extricsic message}),
send $\upsilon_{\eta_{q}^{r}\rightarrow s_{q}^{r}}$ and $\upsilon_{\eta_{q}^{c}\rightarrow s_{q}^{c}}$
to Module B.

\STATE\textbf{\%Module B: Message Passing in MRF}

\STATE Perform message passing, using (\ref{eq:message1}) - (\ref{eq:message5}).

\STATE Calculate the approximate marginal posterior $q\left(\overline{\boldsymbol{s}}\mid\boldsymbol{y};\boldsymbol{\xi}^{\left(t\right)}\right)$,
using (\ref{eq:q(s_bar)}).

\STATE Send $\upsilon_{u_{q}^{r}\rightarrow s_{q}^{r}}$ and $\upsilon_{u_{q}^{c}\rightarrow s_{q}^{c}}$
to Module A.

\STATE \textbf{Turbo-IF-VBI-M Step:}

\STATE Construct the surrogate function $Q\left(\boldsymbol{\xi};\boldsymbol{\xi}^{i}\right)$
in (\ref{eq:func_Q}).

\STATE Update $\boldsymbol{\vartheta}^{i+1}$, using (\ref{eq:update_r})
- (\ref{eq:update_t}).

\STATE Update $\boldsymbol{\zeta}^{i+1}$, using (\ref{eq:update_alpha_2})
and (\ref{eq:update_beta_2}).

\IF{$\left\Vert \boldsymbol{\xi}^{i+1}-\boldsymbol{\xi}^{i}\right\Vert \leq\epsilon$}

\STATE \textbf{break}

\ENDIF

\ENDFOR

\STATE Output $\boldsymbol{\xi}^{\ast}$, $\boldsymbol{x}^{*}=\boldsymbol{\mu}$,
and $s_{i}^{*}=\arg\max_{s_{i}}q\left(s_{i}\mid\boldsymbol{y};\boldsymbol{\xi}^{\ast}\right)$.

\end{algorithmic}
\end{singlespace}
\end{algorithm}

\subsection{Complexity Analysis}

We discuss the computational complexity of the proposed algorithm
and other existing algorithms. The computational complexity of the
Turbo-IF-VBI algorithm is dominated by the update of $q\left(\boldsymbol{v}\mid\boldsymbol{y};\boldsymbol{\xi}\right)$
in Module A. Specifically, the IF-VBI estimator (Module A) involves
some matrix-vector product and diagonal matrix inverse operations
in each iteration. Therefore, the computational complexity order of
the proposed algorithm is $\mathcal{O}\left(M\left(\left|N_{u}\right|+\left|N_{b}\right|\right)\left(1+Q+UV\right)\right)$
per iteration. The existing sparse Bayesian inference algorithms,
such as the Turbo-SBI \cite{Huangzhe_JRC2} and Turbo-VBI \cite{LiuAn_CE_Turbo_VBI},
involve a matrix inverse in each iteration. And the associated computational
complexity order of these algorithms is $\mathcal{O}\left(\left(1+Q+UV\right)^{3}\right)$
per iteration, which is much higher than that of the Turbo-IF-VBI
algorithm.

\section{Simulation Results}

In this section, we evaluate the performance of our proposed scheme
and compare it with some baselines to verify its advantages. The baselines
and our proposed scheme are summarized as follows:
\begin{itemize}
\item \textbf{Baseline 1: }The orthogonal matching pursuit (OMP) algorithm
\cite{Tropp_CE_OMP,Rahman_CS_3D_OMP} with a fixed position grid.
\item \textbf{Baseline 2: }The Turbo-SBI algorithm \cite{Huangzhe_JRC2}.
\item \textbf{Baseline 3:} The Turbo-IF-VBI algorithm with an i.i.d. prior
(the factor graph has no joint support vector $\boldsymbol{\overline{s}}$
and the elements of $\boldsymbol{s}^{r}$ and $\boldsymbol{s}^{c}$
are i.i.d., respectively).
\item \textbf{Proposed:} The Turbo-IF-VBI algorithm with the MRF prior.
\item \textbf{Genie-aided method:} The genie-aided method uses the proposed
scheme based on the assumption that the user location and time offset
are perfectly known.
\item \textbf{Proposed without relaxing:} This scheme is a minor variation
of the proposed scheme. The only different is that Module A directly
maximizes the ELBO without constructing a relaxed ELBO. And thus it
involves a high-dimensional matrix inverse in each iteration.
\end{itemize}
Baseline 1 is a classic compressed sensing algorithm. Baseline 2 is
the state-of-the-art method based on sparse Bayesian inference. Baseline
3 and our proposed scheme use the same algorithm, i.e., the proposed
Turbo-IF-VBI, but with different sparse prior models. The performance
gain between baseline 3 and our proposed scheme reflects the gain
from utilizing the 2-D joint burst sparsity of the location domain
channels.
\begin{figure*}[t]
\centering{}%
\begin{minipage}[t]{0.45\textwidth}%
\begin{center}
\includegraphics[clip,width=80mm]{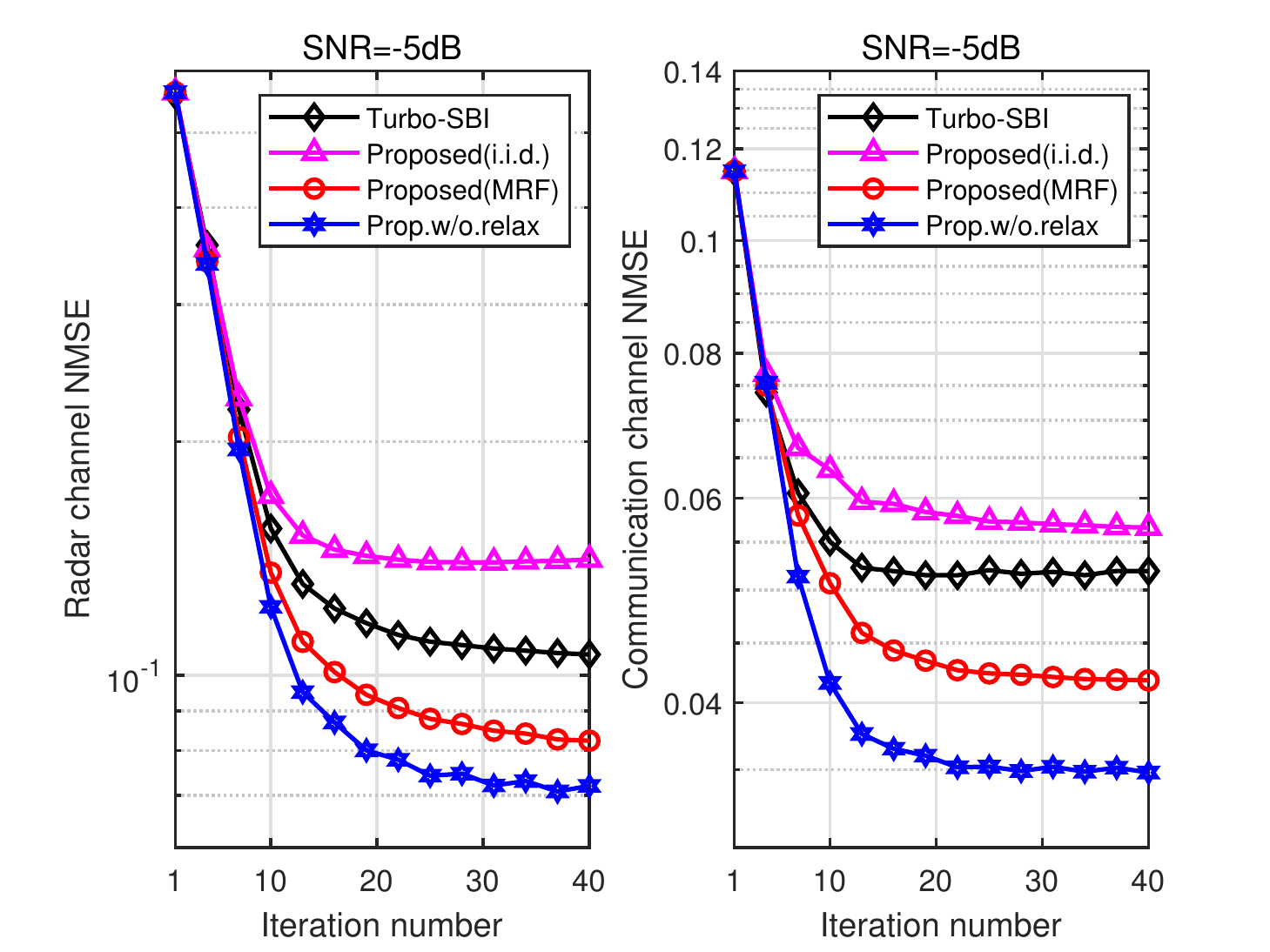}
\par\end{center}
\caption{\label{fig:convergence-behavior}The convergence behavior of different
algorithms when $\textrm{SNR}=-5\ \textrm{dB}$.}
\end{minipage}\hfill{}%
\begin{minipage}[t]{0.45\textwidth}%
\begin{center}
\includegraphics[clip,width=80mm]{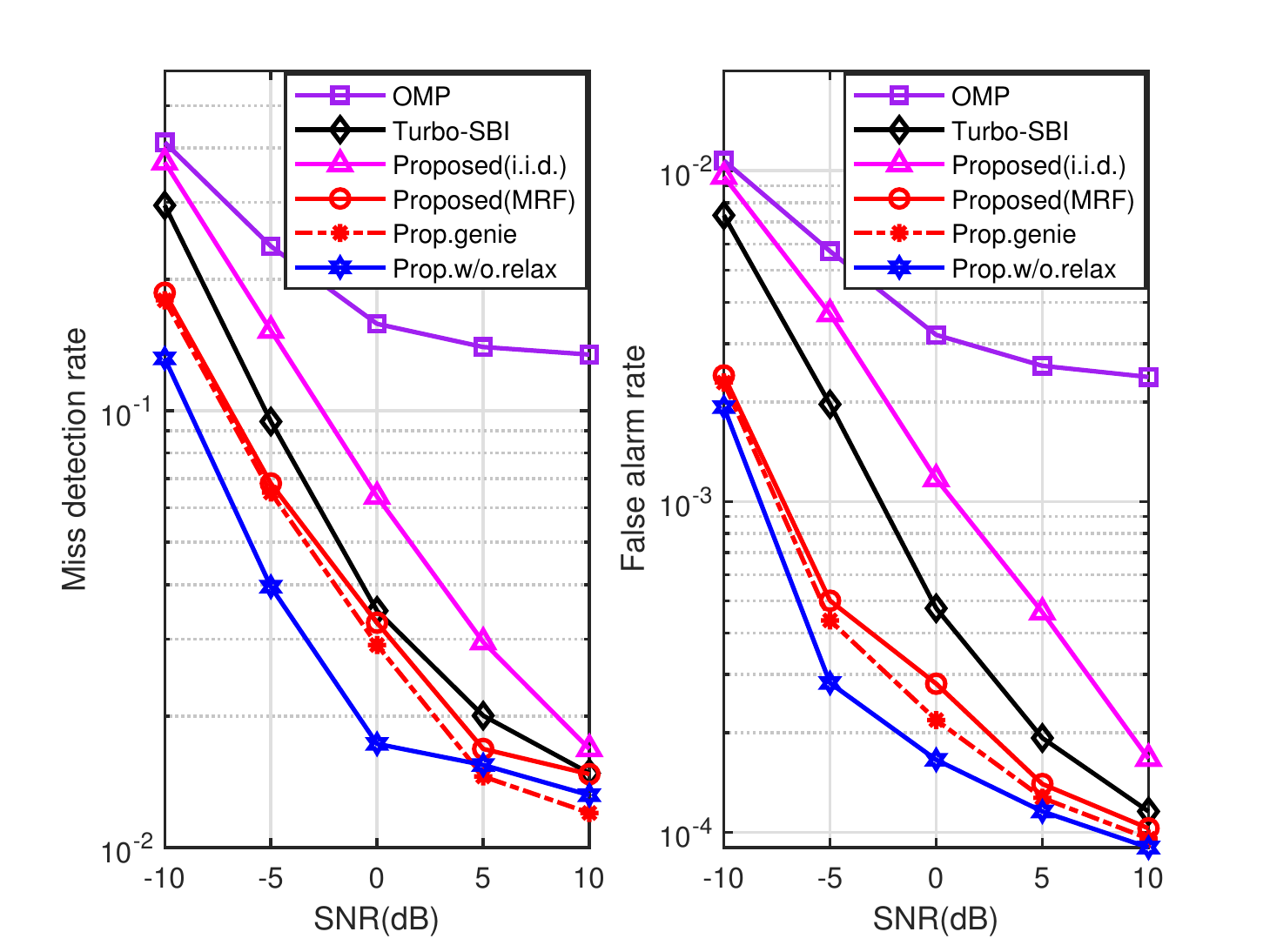}
\par\end{center}
\caption{\textcolor{blue}{\label{fig:MDR-SNR}}Miss detection rate and false
alarm rate of target detection versus SNR.}
\end{minipage}
\end{figure*}
\begin{figure*}[t]
\centering{}%
\begin{minipage}[t]{0.45\textwidth}%
\begin{center}
\includegraphics[clip,width=80mm]{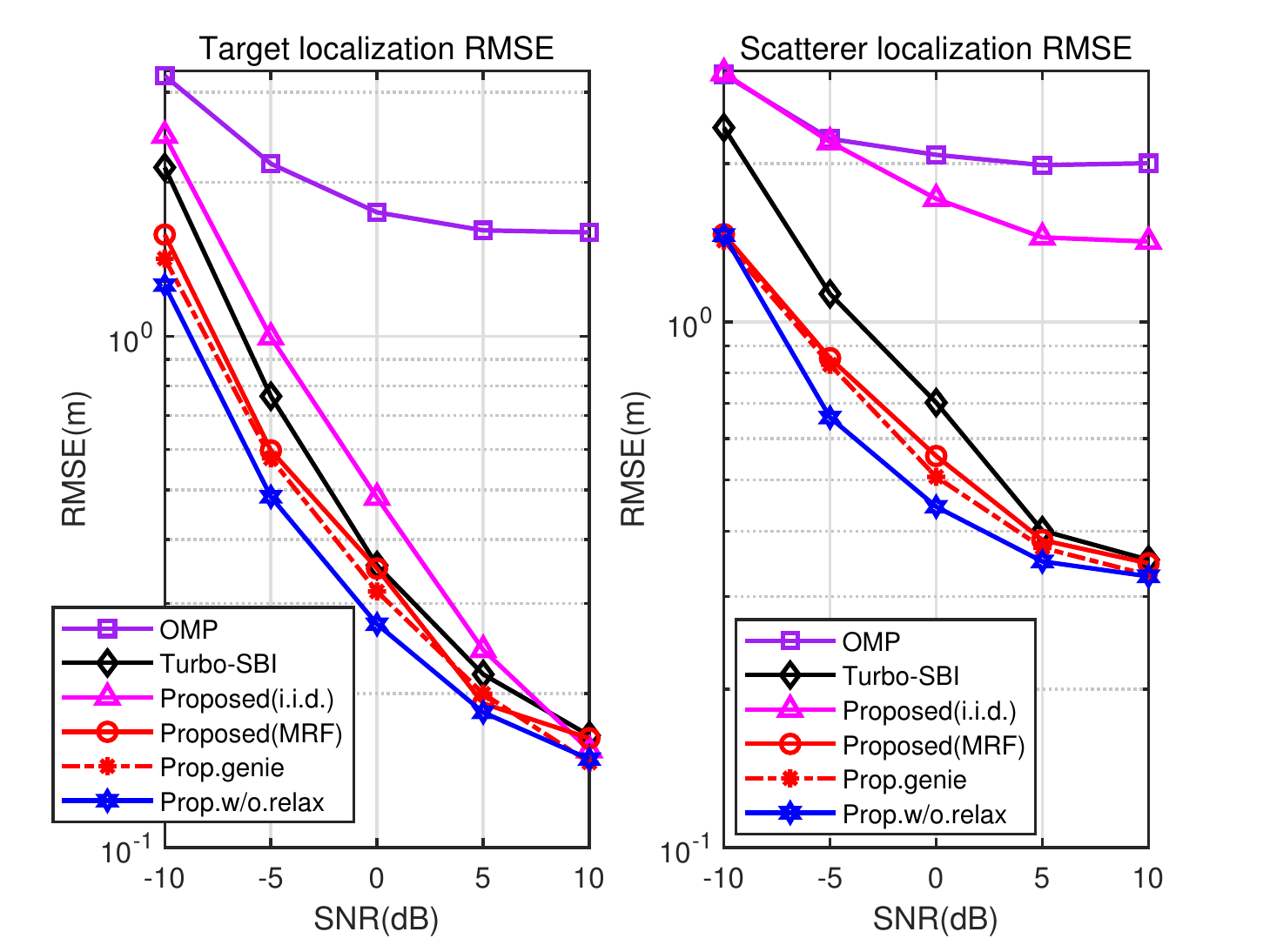}
\par\end{center}
\caption{\label{fig:RMSE-SNR}RMSE of target and scatterer localization versus
SNR.}
\end{minipage}\hfill{}%
\begin{minipage}[t]{0.45\textwidth}%
\begin{center}
\includegraphics[clip,width=80mm]{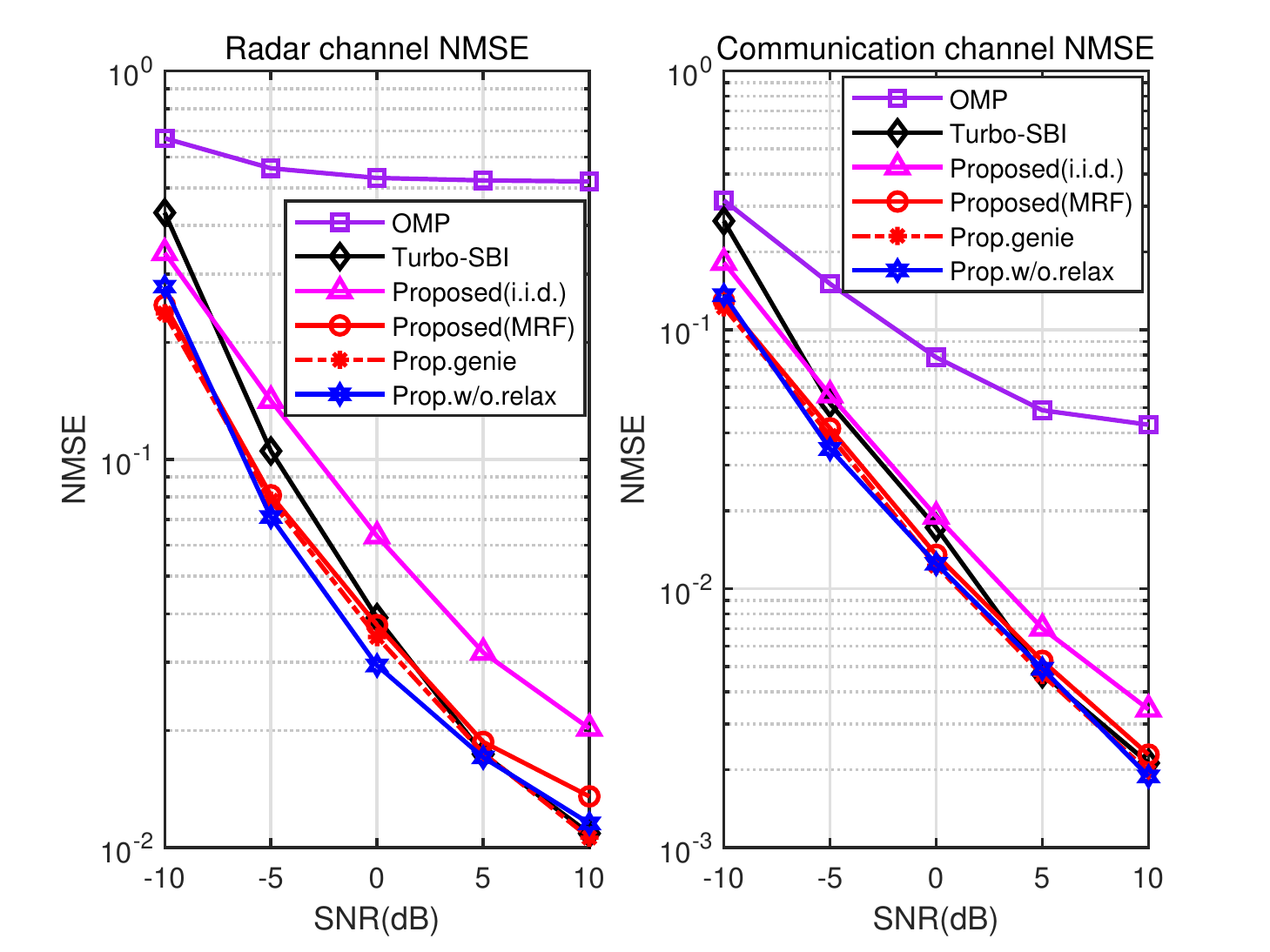}
\par\end{center}
\caption{\textcolor{blue}{\label{fig:NMSE-SNR}}NMSE of radar and communication
channel estimation versus SNR.}
\end{minipage}
\end{figure*}
\begin{figure*}[t]
\centering{}%
\begin{minipage}[t]{0.45\textwidth}%
\begin{center}
\includegraphics[clip,width=80mm]{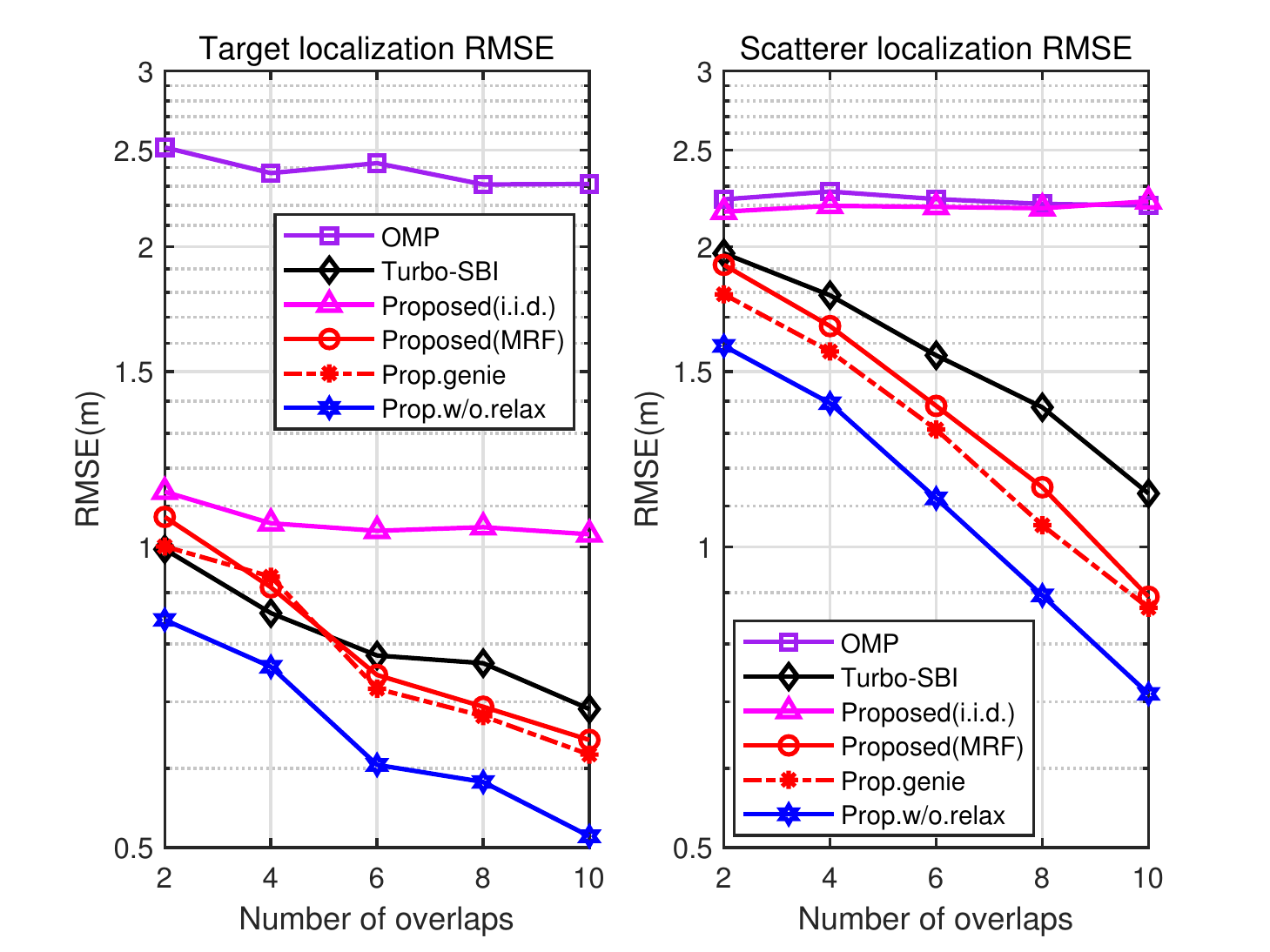}
\par\end{center}
\caption{\label{fig:RMSE-Nu}RMSE of target and scatterer localization versus
number of overlaps.}
\end{minipage}\hfill{}%
\begin{minipage}[t]{0.45\textwidth}%
\begin{center}
\includegraphics[clip,width=80mm]{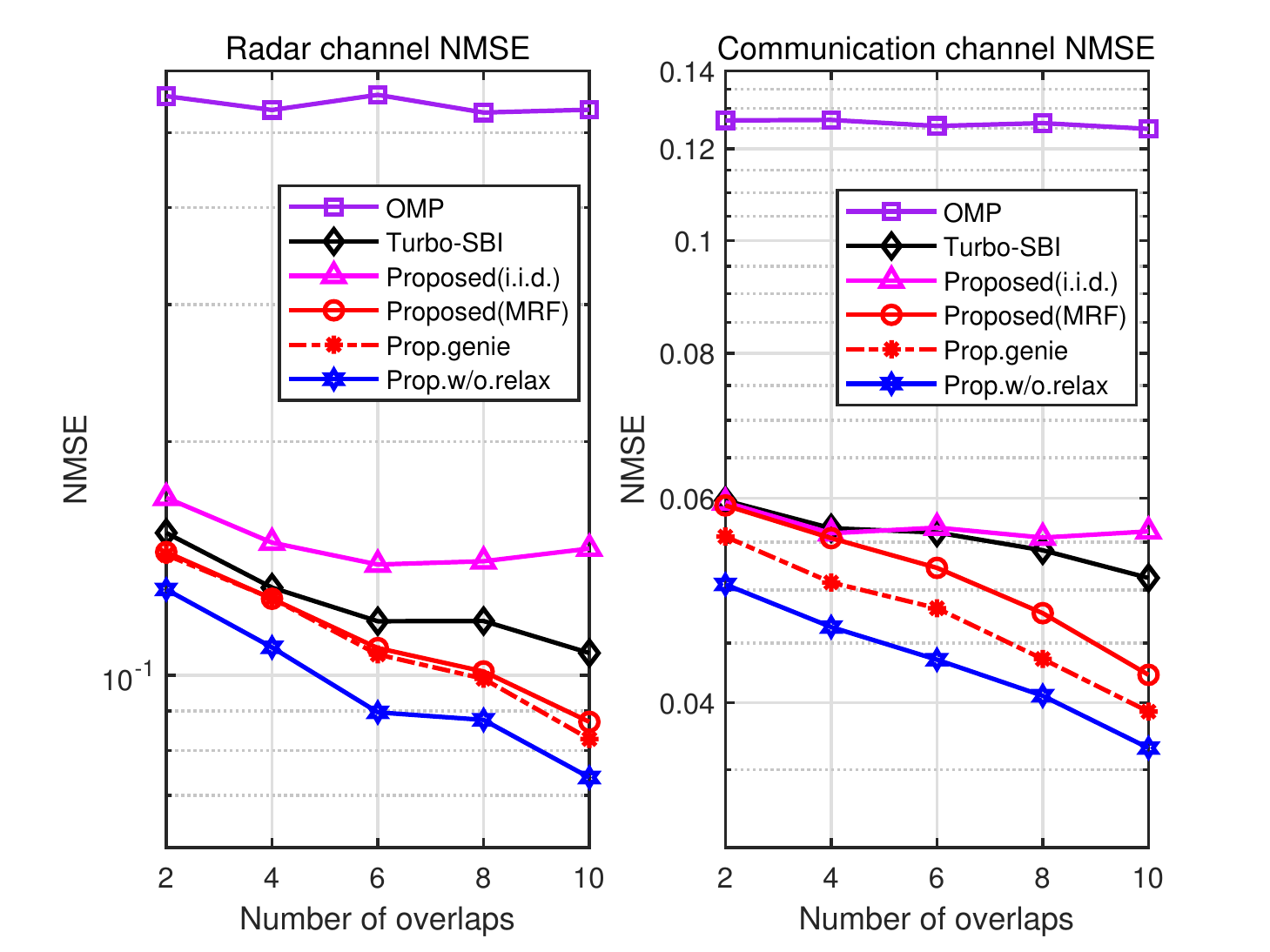}
\par\end{center}
\caption{\textcolor{blue}{\label{fig:NMSE-Nu}}NMSE of radar and communication
channel estimation versus number of overlaps.}
\end{minipage}
\end{figure*}

In the simulations, we consider a $100\text{ m}\times100\text{ m}$
area with a grid resolution of $5\text{ m}$. The BS is at coordinates
$\left[-50\text{ m},0\text{ m}\right]^{T}$ and the mobile user is
around coordinates $\left[50\text{ m},0\text{ m}\right]^{T}$ with
a random position offset. We assume that the prior distribution of
$\boldsymbol{p}_{u}$ is $p^{x}\sim\mathcal{N}\left(50,\sigma_{p}^{2}/2\right)$
and $p^{y}\sim\mathcal{N}\left(0,\sigma_{p}^{2}/2\right)$, where
$\sigma_{p}^{2}$ is set as $1$. There are $K=11$ radar targets
and $L=13$ communication scatterers within the area. Radar targets
and communication scatterers are concentrated on two clusters with
different sizes. The number of OFDM subcarriers is $N=1024$, the
carrier frequency is $3.5\ \textrm{GHz}$, and the subcarrier interval
is $f_{0}=30\text{ kHz}$. Downlink and uplink pilot symbols are generated
with random phase under unit power constrains, and they are inserted
at intervals of $32$ OFDM subcarriers, i.e., $\left|\mathcal{N}_{b}\right|=32$
and $\left|\mathcal{N}_{u}\right|=32$. The BS is equipped with a
ULA of $M=64$ antennas. The time offset $\tau_{o}$ is within $\left[\frac{-2}{B},\frac{2}{B}\right]$,
where $B=Nf_{0}$ denotes the total bandwidth. We use the root mean
square error (RMSE) as the performance metric for target and scatterer
localization and the normalized mean square error (NMSE) as the performance
metric for radar and communication channel estimation. Furthermore,
we also evaluate the target detection performance in terms of miss
detection rate and false alarm rate. To be more specific, $q\left(s_{q}^{r}=1\right)>0.5$
indicates that the BS detects a target lying in the $q\textrm{-th}$
position grid, while $q\left(s_{q}^{r}=1\right)<0.5$ indicates the
opposite, where $q\left(s_{q}^{r}=1\right)$ is the posterior probability
of $s_{q}^{r}=1$ obtained by the algorithms.

\subsection{Convergence Behavior}

Fig. \ref{fig:convergence-behavior} illustrates the convergence behavior
of different algorithms. It can be seen that the proposed Turbo-IF-VBI
algorithm converges quickly within 20 iterations. This further implies
that the approximate marginal posteriors provided by Turbo-IF-VBI-E
Step are accurate enough that they have little effect on the convergence
performance of the whole algorithm. However, the proposed algorithm
with an i.i.d. prior converges to a poor stationary point, while the
proposed algorithm with the MRF prior finds a better solution. It
verifies the advantage of the MRF prior in terms of convergence performance. 

\subsection{Impact of Signal to Noise Ratio (SNR)}

In Fig. \ref{fig:MDR-SNR} - \ref{fig:NMSE-SNR}, we focus on how
the SNR affects sensing/estimation performance. To be more specific,
Fig. \ref{fig:MDR-SNR} shows the miss detection rate and false alarm
rate of target detection, Fig. \ref{fig:RMSE-SNR} shows the RMSE
of target and scatterer localization, and Fig. \ref{fig:NMSE-SNR}
shows the NMSE performance of radar and communication channel estimation.
The performance of all schemes improves as the SNR increases, except
for the OMP. Since the OMP uses a fixed position grid, the performance
of the algorithm is limited by the grid resolution in the high SNR
region. In the low SNR region, the proposed algorithm with the MRF
prior achieves a better performance than the state-of-the-art Turbo-SBI,
which can only exploit the joint burst sparsity in the angle domain.
Besides, the proposed algorithm with the MRF prior gets a significant
performance gain over the proposed algorithm with an i.i.d. prior,
which indicates that the spatially non-stationary MRF model can efficiently
exploit the 2-D joint burst sparsity of the location domain radar
and communication channels. Furthermore, the performance gap between
the genie-aided method and our proposed scheme is very small, which
implies that our proposed scheme can mitigate the impact of the non-ideal
factors effectively. Finally, the scheme without constructing the
relaxed ELBO can achieve the best performance but with higher computational
overhead.

\subsection{Impact of Number of Overlaps}

In Fig. \ref{fig:RMSE-Nu} and Fig. \ref{fig:NMSE-Nu}, we focus on
how the number of overlaps between radar targets and communication
scatterers affects sensing/estimation performance in the case of $\textrm{SNR}=-5\ \textrm{dB}$.
The performance of schemes based on joint estimation improves as the
number of overlaps increases, while the performance of schemes based
on separate estimation (i.e., the proposed algorithm with an i.i.d.
prior and the OMP) remains nearly unchanged. This indicates that the
joint process of radar and communication channels can take advantage
of their correlation to enhance each other's performance. Note that
the scheme without constructing the relaxed ELBO outperforms the genie-aided
method, which reflects that the effect of the matrix inverse approximation
on the performance is slightly larger than that of the imperfect estimation
of the non-ideal factors.

\section{Conclusions}

We propose a joint scattering environment sensing and channel estimation
scheme for a massive MIMO-OFDM ISAC system. A location domain sparse
representation of radar and communication channels is introduced,
which is suitable to the task of joint localization of targets and
scatterers. To capture the 2-D joint burst sparsity of the location
domain channels, we use a spatially non-stationary MRF model that
adapts to different scattering environments that occur in practice.
A Turbo-IF-VBI algorithm is designed, where the E-step uses an inverse-free
algorithm to calculate approximate marginal posteriors of channel
vectors and the M-step applies a low-complexity method to refine the
dynamic position grid, estimate the non-ideal parameters, and learn
the MRF parameters. Simulations verify that our proposed Turbo-IF-VBI
algorithm with the MRF prior achieves a better performance than the
state-of-the-art Turbo-SBI method in \cite{Huangzhe_JRC2}, and meanwhile
avoids the complicated matrix inverse operation in Turbo-SBI.

\bibliographystyle{IEEEtran}
\bibliography{ISAC,Localization_CE,MRF,Others}

\end{document}